\documentclass[%
 reprint,
 amsmath,amssymb,
 aps,
]{revtex4-2}

\usepackage{graphicx}
\usepackage{dcolumn}
\usepackage{bm}
\usepackage{comment}

\begin{document}

\preprint{APS/123-QED}

\title{Light Deflection in Plasma in the Hartle-Thorne Metric and in Other Axisymmetric Spacetimes with a Quadrupole Moment}

\author{Barbora Bezd\v{e}kov\'{a}}
 \email{barbora.bezdekova@mff.cuni.cz}
\author{Ji\v{r}\'{i} Bi\v{c}\'{a}k}%
\affiliation{
Institute of Theoretical Physics, Faculty of Mathematics and Physics, Charles University, V Hole\v{s}ovi\v{c}k\'{a}ch 747/2, Prague 180 00, Czech Republic
}

\date{\today}

\begin{abstract}
The light propagation in plasma medium around stationary gravitating objects is studied in the geometrical optics limit. Static metrics with quadrupole moments are considered, however, the main attention is paid to the stationary Hartle-Thorne metric and Kerr metric. Their different effects on the light deflection are compared. The deflection angle is given in a general analytic form, the detailed calculations are performed in the weak field limits and illustrated graphically. The trajectories of the light rays are constructed and compared, emphasizing the role of the quadrupole moment.
\end{abstract}

\maketitle


\section{Introduction}
Stable compact astrophysical objects, after having been formed by gravitational collapse are likely to be stationary and axisymmetric. If the initial conditions are such that the collapse leads to a black hole characterized just by its mass $M$ and angular momentum $J$, the resulting spacetime will be described by a Kerr metric. (We assume a possible small electric charge can be neglected.) A Kerr black hole has specific multipole moments, all are expressible in terms of $M$ and $J$. However, the Universe is populated by a number of other compact rotating objects like neutron stars, many of them observed as pulsars or X-ray sources. Possibly, more exotic compact objects like rotating boson stars or wormholes exist.

In 1911, Einstein published the most often quoted paper written during his Prague stay, in which, on the basis of the preliminary version of his gravity theory, he suggested that a light ray propagating near Sun will be deflected by an observable amount due to Sun's gravitational field.  After the complete Einstein's general relativity was formulated in November 1915, the effect of 1.75 arcseconds deflection was predicted (twice as large as was the Prague prediction). In 1919 it was famously confirmed by expeditions led by Eddington and Dyson (see, e.g., Ref.~\cite{Renn}).  \citet{EHTC19}, 100 years later, released an image of the black hole in the center of galaxy M87 which revealed how rich are effects of the strong lensing around a black hole. Numerous papers were published since then, over 250 of them are quoted in a very comprehensive recent review \cite{PerlickTsupko22} on analytical studies of black hole shadows.
Just a few works, however, consider, together with the effects of black-hole field, the influence of a plasma surrounding the hole on the behavior of light rays. For an example, see Ref.~[\onlinecite{liu17}] where the deflection angle in Kerr metric in homogeneous plasma in terms of a strong gravitational lensing is analyzed.

The main purpose of our work is to study the deflection angle of rays propagating in a plasma around rotating, gravitating objects with a general \emph{quadrupole moment} and make comparison with similar situations in the Kerr case with plasma. We concentrate primarily on the Hartle-Thorne (HT) metrics (cf Ref.~\cite{HartleThorne1968}) representing slowly rotating stars with quadrupole moments. We define the multiple moments in accordance with basic paper by \citet{geroch70}. We include effects of surrounding plasma on the rays. In Ref.~\cite{Berti2005} it is analyzed in detail that ``the Hartle-Thorne approximation is very reliable for most astrophysical applications''... The authors  integrate the HT structure equations for five equations of state and match the models to numerical solutions of the full Einstein equations assuming the same mass and the same angular moment. They compare the Hartle-Thorne approximation with numerical solutions of the full nonlinear Einstein's equations and show that deviations in the quadrupole moment $Q$ are not more than 20\% for the fastest observed pulsar PSR J1939+2134.

In Section 2 we find the deflection angle formula for a general axially symmetric stationary metric. We start from an elegant Hamiltonian formalism introduced in Ref.~\cite{Synge} for the light propagation in a general spacetime with a (locally isotropic) refractive and dispersive medium. We assume the medium to be stationary (with its 4-velocity pointing along the timelike Killing vector of the spacetime), and we assume the rays are moving in the equatorial plane.

In Section 3 the deflection angle is calculated for the Hartle-Thorne metric accurate to the second order in the angular velocity. The deflection angle formula for the Kerr metric is first derived in the Boyer-Lindquist coordinates (Section 3.2) and then compared with the expression for the deflection angle for the HT metric in the weak field approximation (Section 3.3). Some more cumbersome calculations of the terms entering the final expressions for both the Kerr and HT metrics in the weak-field limit are relegated to Appendices~\ref{App_HT}, \ref{App_Kerr}, and \ref{App_HT_weak}.

Some other metrics with a quadrupole moment (e.g., Erez-Rosen metric or so-called $q$-metric) are briefly presented in Section 4.

In Section 5 there are number of figures constructed illustrating the dependence of some exact and approximate deflection angles on the minimal distance $R/M$ the rays can reach from the source which is either surrounded by vacuum or by a plasma medium. For an explicit illustration, the ray trajectories are plotted in both vacuum and plasma case analyzed in the preceding Sections.

The discussion of the results and some final remarks follow in the concluding Section 6.

\section{Deflection Angle in an Axially Symmetric Stationary Spacetime}
An axially symmetric stationary spacetime can be described by the metric
\begin{align}\label{metrika}
ds^2 =&-A(r,\vartheta)dt^2  + B(r,\vartheta)dr^2+2P(r,\vartheta) dtd\varphi \\
&+D(r,\vartheta)d\vartheta^2
+C(r,\vartheta) d\varphi^2,\nonumber
\end{align}
where $A(r,\vartheta),B(r,\vartheta),C(r,\vartheta),D(r,\vartheta)$, and $P(r,\vartheta)$ are functions of radial coordinate $r$ and polar coordinate $\vartheta$. To guarantee that the Killing vector fields, $\partial/\partial t$ and $\partial/\partial \varphi$, span timelike surfaces, it must hold
that $A(r,\vartheta)C(r,\vartheta) + P(r,\vartheta)^2 > 0$, $B(r,\vartheta) > 0$, and $D(r,\vartheta) > 0$ (for the details see Ref.~\cite{bezdekova22}). The functions can generally depend on other parameters, for example, on specific angular momentum $a$. These are constant factors not relevant in the derivation of equations of motion given below.

The inverse metric to (\ref{metrika}) is
\begin{align}
  g^{rr} &= \frac{1}{B(r,\vartheta)},\quad g^{\vartheta \vartheta} = \frac{1}{D(r,\vartheta)}, \quad \\
  g^{\varphi \varphi} &= \frac{A(r,\vartheta)}{A(r,\vartheta)C(r,\vartheta)+P^2(r,\vartheta)}, \nonumber\\
  g^{tt} &= \frac{-C(r,\vartheta)}{A(r,\vartheta)C(r,\vartheta)+P^2(r,\vartheta)}, \quad \nonumber\\
  g^{t\varphi} &= \frac{P(r,\vartheta)}{A(r,\vartheta)C(r,\vartheta) + P^2(r,\vartheta)}.\nonumber
\end{align}

Considering the ray propagation through a refractive, dispersive medium characterized by refractive index $n$ and 4-velocity $V^\alpha$, a corresponding Hamiltonian takes the form \cite{Synge}
\begin{equation}\label{hamiltonian}
\mathcal{H}(x^{\alpha},p_{\alpha})=\frac{1}{2}\left[g^{\beta\delta}p_{\beta}p_{\delta}-(n^2-1)(p_{\gamma}V^{\gamma})^2\right].
\end{equation}
If we further assume the medium to be at rest in our coordinate frame (see Ref.~\cite{Synge}), we have \footnote{We assume $i,j=1,2,3$; $\alpha, \beta=0,1,2,3$.} $V^{i}=0$, $V^{t}=(-g_{tt})^{-1/2}$ and the Hamiltonian can be rewritten as
\begin{equation}
\mathcal{H}(x^{\alpha},p_{\alpha})=\frac{1}{2}\left[g^{\beta\delta}p_{\beta}p_{\delta}+(n^2-1)p_t^2g_{tt}^{-1}\right].
\end{equation}
Moreover, due to the dependence of photon frequency $\omega(x^{\alpha})$ as measured by an observer at rest in the medium at point $x^{\alpha}$, we get 
\begin{equation}\label{omega_alpha}
\omega(x^{\alpha})=-p_{\alpha}V^{\alpha},
\end{equation}
so $\omega(x^{\alpha})=-p_{t}(-g_{tt})^{-1/2}$. In a general dispersive medium $n=n(x^\alpha,\omega(x^{\alpha}))$.
From the relations above, we can obtain
\begin{align}\label{omega}
\omega(r)=-p_{t}\frac{1}{\sqrt{A(r,\vartheta)}}.
\end{align}

With metric (\ref{metrika}), one finds
\begin{widetext}
\begin{gather}\label{hamiltonian_conc}
\mathcal{H}(x^\alpha,p_{\alpha})=\frac{1}{2}\left[\frac{p_{r}^2}{B(r,\vartheta)}+\frac{p_{\vartheta}^2}{D(r,\vartheta)}+ \frac{p_{\varphi}^2A(r,\vartheta)}{A(r,\vartheta)C(r,\vartheta)+P^2(r,\vartheta)} -\frac{p_{t}^2C(r,\vartheta)}{A(r,\vartheta)C(r,\vartheta)+P^2(r,\vartheta)} \right.\\
\left.+\frac{2p_tp_{\varphi}P(r,\vartheta)}{A(r,\vartheta)C(r,\vartheta)+P^2(r,\vartheta)}+\frac{p_{t}^2(1-n^2)}{A(r,\vartheta)}\right].\nonumber
\end{gather}
\end{widetext}

Since for an asymptotically flat spacetime $g^{tt} \rightarrow -1$ at $r \rightarrow \infty$ in adapted coordinates, we get $\omega(\infty)=-p_t\equiv\omega_0$. Furthermore, we assume that medium has refractive index $n=n(r,\omega(r))$.

Using the Hamiltonian~(\ref{hamiltonian_conc}), we get the equations of motion in the form
\begin{align}
  \dot{\varphi} &= \frac{\partial \mathcal{H}}{\partial p_{\varphi}}=\frac{p_{\varphi}A(r,\vartheta)-P(r,\vartheta)\omega_0}{A(r,\vartheta)C(r,\vartheta)+P^2(r,\vartheta)}, \label{eom1}\\
  \dot{r} &= \frac{\partial \mathcal{H}}{\partial p_r}=\frac{p_{r}}{B(r,\vartheta)}.\label{eom2}
\end{align}

From $\mathcal{H}(x^\alpha,p_{\alpha})=0$, it is possible to obtain the expression for $p_r$. It reads
\begin{widetext}
\begin{align}\label{pr}
p_r=\pm \sqrt{B(r,\vartheta)}\sqrt{\frac{\omega_{0}^2C(r,\vartheta)+2\omega_0p_{\varphi}P(r,\vartheta)-p_{\varphi}^2A(r,\vartheta)}{A(r,\vartheta)C(r,\vartheta)+P^2(r,\vartheta)}-\frac{\omega_{0}^2(1-n^2)}{A(r,\vartheta)}-\frac{p_{\vartheta}^2}{D(r,\vartheta)}}.
\end{align}
\end{widetext}

The relations given above allow us to derive the following equation of motion for the rays:
\begin{equation}
\frac{d\varphi}{dr}=\frac{\dot{\varphi}}{\dot{r}}=\frac{B(r,\vartheta)}{p_{r}}\frac{p_{\varphi}A(r,\vartheta)-P(r,\vartheta)\omega_0}{A(r,\vartheta)C(r,\vartheta)+P^2(r,\vartheta)}.
\end{equation}
Let us further assume that the rays are moving in the equatorial plane, so $\vartheta=\pi/2$ and hence $p_\vartheta=0$. This assumption enables one to obtain a total deflection angle in analytical form. The equatorial plane is the most natural choice, also regarding the well-known Bardeen-Petterson effect.
Using the expression (\ref{pr}) for $p_r$ implies
\begin{widetext}
\begin{gather}\label{poh_rovnice}
\frac{d\varphi}{dr}=\pm \frac{\sqrt{A(r)B(r)}\left(\frac{p_{\varphi}}{\omega_0}-\frac{P(r)}{A(r)}\right)}{\sqrt{A(r)C(r)+P^2(r)}}
\left(\frac{C(r)}{A(r)}-\frac{p_{\varphi}^2}{\omega_{0}^2}+2\frac{p_{\varphi}}{\omega_0}\frac{P(r)}{A(r)}-(1-n^2)\frac{A(r)C(r)+P^2(r)}{A^2(r)}\right)^{-1/2}.
\end{gather}
\end{widetext}
In fact, the ray trajectory consists of two parts - where the $r$ coordinate either decreases or increases along with $\varphi$ increasing. In each part of the motion corresponding sign ($+$ for $r$ increasing, $-$ for $r$ decreasing) in equation (\ref{poh_rovnice}) is used.

Formula (\ref{poh_rovnice}) can be further rearranged as
\begin{align}
\frac{d\varphi}{dr}&=\pm \sqrt{\frac{A(r)B(r)}{A(r)C(r)+P^2(r)}}\\
&\times\left(\frac{n^2\left(\frac{C(r)}{A(r)}+\frac{P^2(r)}{A^2(r)}\right)}{\left(\frac{p_{\varphi}}{\omega_{0}}-\frac{P(r)}{A(r)}\right)^2}-1
\right)^{-1/2}.\nonumber
\end{align}

Similarly to the previous work (e.g., Ref.~\cite{PerlickTsupko22}), it is useful to introduce function $h^2(r)$ defined as
\begin{equation}
h^2(r)=n^2\left(\frac{C(r)}{A(r)}+\frac{P^2(r)}{A^2(r)}\right).
\end{equation}
For refractive index $n$ in a cold plasma approximation it holds
\begin{equation}\label{def_ref_index}
n^2=1-\frac{\omega_{p}^2(r)}{\omega^2(r)}=1-\frac{\omega_{p}^2(r)}{\omega_0^2}A(r),
\end{equation}
because the photon frequency $\omega(x)$ can be obtained from the relation (\ref{omega_alpha}); plasma frequency $\omega_{p}(r)$ is given by the properties of the medium surrounding the gravitating object. (Formula (\ref{def_ref_index}) holds only in an unmagnetized cold plasma, while in a more general case, additional dispersion relation would arise, and a non-isotropic tensor would be needed to describe the medium.) Hence, function $h^2(r)$ can be rewritten as
\begin{equation}
h^2(r)=\frac{A(r)C(r)+P^2(r)}{A^2(r)}\left(1-\frac{\omega_{p}^2(r)}{\omega_{0}^2}A(r)\right).
\end{equation}

The deflection angle formula in an axially symmetric stationary spacetime with a refractive medium can thus be written in a general form as follows:
\begin{align}
\alpha=&\pm 2\int_{R}^{\infty}\sqrt{\frac{A(r)B(r)}{A(r)C(r)+P^2(r)}} \\
&\times\left(\frac{h^2(r)}{\left(\frac{p_{\varphi}}{\omega_{0}}-\frac{P(r)}{A(r)}\right)^2}-1\right)^{-1/2}dr - \pi.\nonumber
\end{align}
Here $R$ is the turning point of the ray trajectory, i.e., the minimal value that $r$ coordinate can reach. At this point it hence holds
\begin{equation}
 \left.\frac{dr}{d\varphi}\right|_{r=R}=0.
\end{equation}
This condition is satisfied when
\begin{equation}\label{def_hR}
h^2(R)=\left(\frac{p_{\varphi}}{\omega_{0}}-\frac{P(R)}{A(R)}\right)^2.
\end{equation}
Since $\frac{p_{\varphi}}{\omega_{0}}$ defines the commonly used impact parameter $b$, formula (\ref{def_hR}) actually shows a relation between function $h^2(R)$ and $b$. In this case, the impact parameter can thus be expressed as
\begin{equation}\label{def_b}
b=\frac{p_{\varphi}}{\omega_{0}}=\frac{P(R)}{A(R)}\pm h(R)=\frac{P(R)}{A(R)}\pm n\sqrt{\frac{C(R)}{A(R)}+\frac{P^2(R)}{A^2(R)}}.
\end{equation}
When definition (\ref{def_hR}) is applied, the deflection angle formula finally takes the form
\begin{align}\label{ohyb_uhel}
\alpha&=\pm 2\int_{R}^{\infty}\sqrt{\frac{A(r)B(r)}{A(r)C(r)+P^2(r)}} \\
&\times\left(\frac{h^2(r)}{\left(\frac{P(R)}{A(R)}-\frac{P(r)}{A(r)}\pm h(R)\right)^2}-1\right)^{-1/2}dr - \pi.\nonumber
\end{align}
Due to the relation (\ref{def_b}), the deflection angle can also be expressed as a function of $b$. The deflection angles as a function of impact parameter in the case of the metrics discussed in the present work are derived in Appendix~\ref{App_b}.

\section{Deflection Angle in the Hartle-Thorne Metric}
\subsection{Relevant terms of the Hartle-Thorne Metric}
Let us now employ formula (\ref{ohyb_uhel}) for the derivation of the deflection angle for the Hartle-Thorne (HT) metric \cite{HartleThorne1968} and compare the results with the deflection angle in the Kerr metric.

The external gravitational field of the rotating star, accurate to the second order in the angular velocity, takes the form
\begin{widetext}
\begin{align} \label{HT_metrika}
ds^2 =-&\left(1-\frac{2M}{r}+\frac{2J^2}{r^4}\right)\left\{1+2P_2(\cos\vartheta)\left[\frac{J^2}{Mr^3}\left(1+\frac{M}{r}\right)+\frac{5}{8}\frac{Q-J^2/M}{M^3}Q_{2}^2\left(\frac{r}{M}-1\right)\right]\right\}dt^2\\
&+\left(1-\frac{2M}{r}+\frac{2J^2}{r^4}\right)^{-1}\left\{1-2P_2(\cos\vartheta)\left[\frac{J^2}{Mr^3}\left(1-\frac{5M}{r}\right)+\frac{5}{8}\frac{Q-J^2/M}{M^3}Q_{2}^2\left(\frac{r}{M}-1\right)\right]\right\}dr^2\nonumber\\
+r^2&\left\{1+2P_2(\cos\vartheta)\left[-\frac{J^2}{Mr^3}\right.\right.\left.\left.\left(1+\frac{2M}{r}\right)+\frac{5}{8}\frac{Q-J^2/M}{M^3}\left\langle\frac{2M}{\sqrt{r(r-2M)}}Q_{2}^1\left(\frac{r}{M}-1\right)-Q_{2}^2\left(\frac{r}{M}-1\right)\right\rangle\right]\right\} \nonumber\\
&\times\left\{d\vartheta^2+\sin^{2}\vartheta \left(d\varphi-\frac{2J}{r^3}dt\right)^2\right\},\nonumber
\end{align}
\end{widetext}
where $M$ is the total mass, $J$ is the total angular momentum of the star, $Q$ is the quadrupole moment, and $Q_{2}^1(x)$, $Q_{2}^2(x)$ are the associated Legendre functions of the second kind, and $P_2(\cos\vartheta)$ represents the Legendre polynomial of the second order. Let us introduce compact notation for the following dimensionless quantities:
\begin{align}\label{compact_notation}
  A_1 &= 1-\frac{2M}{r}+\frac{2J^2}{r^4}, \quad &j =& \frac{J^2}{Mr^3}, \\
  K &= \frac{5}{8}\frac{Q-J^2/M}{M^3}, \quad &j_1 =& \frac{2J}{r^2}, \nonumber\\
  Q_{2}^1&=Q_{2}^1\left(\frac{r}{M}-1\right), \quad &Q_{2}^2=&Q_{2}^2\left(\frac{r}{M}-1\right).\nonumber
\end{align}
We again restrict the motion to the equatorial plane, so $\vartheta=\pi/2$ and $P_2(\cos\vartheta)|_{\vartheta=\frac{\pi}{2}}=-\frac{1}{2}$. The terms relevant for the deflection angle can then be calculated explicitly. They are derived in Appendix~\ref{App_HT}.

We now start from the expressions (\ref{HT_equa1})-(\ref{HT_equa2}) for the deflection angle given in Appendix~\ref{App_HT} and keep terms up to the order $\mathcal{O}(\frac{J^2}{M^2R^2})$ and the lowest nonzero terms $\sim K$. The terms entering the deflection angle formula (\ref{ohyb_uhel}) then read as follows:
\begin{alignat}{2}
 &A(r)B(r) = 1-j\frac{6M}{r}-j_{1}^2\left(1-\frac{2M}{r}\right)^{-1}\times\\
 &\left[1- K\left(\frac{2M}{\sqrt{r(r-2M)}}Q_{2}^1-Q_{2}^2\right)\right]-2jKQ_{2}^2\left(1-\frac{2M}{r}\right),\nonumber\\
 &A(r)C(r)+P^2(r)=A_1r^2+jr^2\frac{M}{r}\left(1-\frac{2M}{r}\right)\\
 &-KA_1r^2\frac{2M}{\sqrt{r(r-2M)}}Q_{2}^1-jKr^2\left(1-\frac{2M}{r}\right)\nonumber\\
&\times\left[\left(2+\frac{3M}{r}\right)Q_{2}^2-\left(1+\frac{M}{r}\right)\frac{2M}{\sqrt{r(r-2M)}}Q_{2}^1\right],\nonumber\\
&1-\frac{\omega_{p}^2(r)}{\omega_{0}^2}A(r)=1-\frac{\omega_{p}^2(r)}{\omega_{0}^2}\left[A_1-j\left(1-\frac{2M}{r}\right)\left(1+\frac{M}{r}\right)\right.\\
&\left.-KA_1Q_{2}^2-j_{1}^2+j_{1}^2K\left(\frac{2M}{\sqrt{r(r-2M)}}Q_{2}^1-Q_{2}^2\right)\right],\nonumber\\
&\frac{P(r)}{A(r)}=-rj_1\left(1-\frac{2M}{r}\right)^{-1}\\
&\times\left[1-K\left(\frac{2M}{\sqrt{r(r-2M)}}Q_{2}^1-2Q_{2}^2\right)\right].\nonumber
\end{alignat}

Now, let us compare the deflection angle formula for the Kerr metric with that for the HT metric up to $\mathcal{O}\left(\frac{M^3}{R^3}\right)$. For this reason, the expansion of the Kerr metric including terms of the order $\frac{M}{R}$, $\frac{a}{R}$, $\frac{M^2}{R^2}$, $\frac{Ma}{R^2}$, $\frac{a^2}{R^2}$, $\frac{M^3}{R^3}$, $\frac{M^2a}{R^3}$, $\frac{Ma^2}{R^3}$, and $\frac{a^3}{R^3}$ is needed. The reasons for the choice of the third order expansion will be clarified further. Before that, the expansion of the deflection angle in the Kerr metric in the Boyer-Lindquist coordinates is derived. To compare both the HT and Kerr metrics, one has to set $J=-Ma$ and $Q=J^2/M$ in the HT metric (see \cite{HartleThorne1968}).

\subsection{Deflection angle formula in the Kerr metric in the Boyer-Lindquist coordinates}
The deflection angle formula in the Kerr spacetime with plasma in the Boyer-Lindquist coordinates in the equatorial plane reads (e.g., Ref.~\cite{perlick2000})
\begin{align}
&\alpha=\pm 2\int_{R}^{\infty}\frac{\sqrt{r(r-2M)}}{r^2-2Mr+ a^2}\\
&\times\left(\frac{(R-2M)^2h^2(r)}{\left(2Ma(R-r)\pm (r-2M)h(R)\right)^2}-1\right)^{-1/2}dr -\pi,\nonumber
\end{align}
where
\begin{equation}
h^2(r)=r^2(r^2-2Mr+a^2)\left(1-\frac{\omega_{p}^2(r)}{\omega_{0}^2}\left(1-\frac{2M}{r}\right)\right)
\end{equation}
and similarly for $h^2(R)$.
This result can be derived from our general formula (\ref{ohyb_uhel}) when considering that
\begin{gather}
 A(r) = 1-\frac{2M}{r}, \quad B(r)= \frac{r^2}{r^2 -2Mr+ a^2},\\
 C(r)= r^2 + a^2 +\frac{2Ma^2}{r}, \quad P(r) = -\frac{2Ma}{r}.\nonumber
\end{gather}
Notice that the terms are given in the equatorial plane and they are hence simplified.

The form of the deflection angle in the Kerr metric in the Boyer-Lindquist coordinates in the weak field approximation (when $M/R\ll 1$) in vacuum is then
\begin{align}\label{alpha_BL}
\alpha_{BL}&=\frac{4M}{R}+\frac{M^2}{R^2}\left(\frac{15\pi}{4}-4\right) \mp\frac{4Ma}{R^2}\\
&\mp\frac{M^2a}{R^3}(10\pi-16)+\frac{2Ma^2}{R^3}+\frac{M^3}{R^3}\left(\frac{122}{3}-\frac{15\pi}{2}\right).\nonumber
\end{align}
Terms up to the second order were calculated in Ref.~\cite{PerlickTsupkoNN} and a complete derivation of formula (\ref{alpha_BL}) can be found in Appendix~\ref{App_Kerr}. The deflection angle formula in the weak field approximation in the Schwarzschild metric follows from (\ref{alpha_BL}) when setting $a=0$. The first term in (\ref{alpha_BL}) is the so-called Einstein angle (see Ref.~\cite{schneider92,bisnovatyi09}). Analogous expressions for the Kerr metric have been obtained by different techniques in several other previous studies, e.g., Ref.~\cite{iyer09,aazami11,crisnejo19}. In those works, the deflection angle was derived in terms of mass~$M$, angular momentum per unit mass~$a$, and impact parameter~$b$ (cf. Appendix~\ref{App_b}).
However, when one wants \emph{to compare the deflection angles of the Kerr and HT metric}, it is necessary to transform the Kerr metric into the appropriate coordinates. The transformation originally introduced by \citet{HartleThorne1968} reads
\begin{gather}
r \rightarrow  r\left[1-\frac{a^2}{2r^2}\left(\left(1+\frac{2M}{r}\right)\left(1-\frac{M}{r}\right)\right.\right.\\
\left.\left.-\cos^2\vartheta\left(1-\frac{2M}{r}\right)\left(1+\frac{3M}{r}\right)\right)\right],\nonumber\\
\vartheta \rightarrow \vartheta -\frac{a^2}{2r^2}\sin\vartheta\cos\vartheta\left(1+\frac{2M}{r}\right).\nonumber
\end{gather}
In the equatorial plane the relevant form of the transformation up to the given order is
\begin{gather}\label{transf_Kerr}
r \rightarrow r\left[1-\frac{a^2}{2r^2}\left(1+\frac{M}{r}\right)\right], \quad \vartheta \rightarrow \vartheta.
\end{gather}

The transformation implies only the change in the term $\propto \frac{Ma^2}{R^3}$. We now obtain \footnote{Relevant transformation factors occur in \begin{gather*}
 \frac{\sqrt{r(r-2M)}}{r^2-2Mr+a^2}\approx\frac{1}{r}\left(1+\frac{M}{r}-\frac{a^2}{r^2}+\frac{3M^2}{2r^2}+\frac{5M^3}{2r^3}-\frac{3Ma^2}{r^3}\right)\\
  \rightarrow \quad \frac{1}{r}\left[1+\frac{a^2}{2r^2}\left(1+\frac{M}{r}\right)\right]\\
  \times\left[1+\frac{M}{r}\left(1-\frac{a^2}{2r^2}\right)-\frac{a^2}{r^2}+\frac{3M^2}{2r^2}+\frac{5M^3}{2r^3}-\frac{3Ma^2}{r^3}\right].
\end{gather*}
Factors arising from these transformations in the other terms vanish. For more details about the integration, see Appendix~\ref{App_Kerr}.}
\begin{align}
  \int_{R}^{\infty} \frac{3Ma^2(r^4+R(R+r)(r^2-R^2))dr}{2r^4R^2(r+R)\sqrt{r^2-R^2}}&=\frac{2Ma^2}{R^3}.
\end{align}

Hence, in the new coordinate system the deflection angle formula for the Kerr metric with $M/R\ll1$ becomes 
\begin{align}\label{alpha_Kerr_transf}
&\alpha_{transf}=\frac{4M}{R}+\frac{M^2}{R^2}\left(\frac{15\pi}{4}-4\right) \mp\frac{4Ma}{R^2}\\
&\mp\frac{M^2a}{R^3}(10\pi-16)+\frac{4Ma^2}{R^3}+\frac{M^3}{R^3}\left(\frac{122}{3}-\frac{15\pi}{2}\right).\nonumber
\end{align}

\subsection{Deflection angle formula in the HT metric in the weak field approximation}\label{Sec_HT_weak}
To be able to directly compare the terms of the deflection angle for the Kerr metric and for the HT metric, we express the relevant HT metric terms obtained above up to the third order in $\frac{M}{R}$, maintaining terms up to the second order in angular velocity 
when the approximation of HT metric as an external gravitational field of the rotating star is still valid. In this order, it is also possible to see the effect of the quadrupole moment. The HT metric as an approximation of the external field of nonrelativistic stars was already given by \citet{HartleThorne1968}. We aim to compare the deflection angles in plasma in both the HT and Kerr metric in the weak field. For this reason, it is useful to derive the HT metric in the weak field approximation directly. At first, we compare the deflection angle in the HT and Kerr metric in vacuum without a quadrupole moment. In the equatorial plane and when $K=0$ (cf. (\ref{compact_notation})) which is valid in the Kerr case, the corresponding terms take the form
\begin{align}
&A(r)= 1-\frac{2M}{r}-j\equiv A_{0}, \\
&A(r)B(r)=1\equiv AB_{0},\\
&A(r)C(r)+P^2(r)=r^2\left(1-\frac{2M}{r}\right)\equiv ACP_{0},\\
&1-\frac{\omega_{p}^2(r)}{\omega_{0}^2}A(r)=1-\frac{\omega_{p}^2(r)}{\omega_{0}^2}\left(1-\frac{2M}{r}-j\right)\\
&\equiv 1-\frac{\omega_{p}^2(r)}{\omega_{0}^2}A_{0},\nonumber\\
&\frac{P(r)}{A(r)}=-\frac{2J}{r}\left(1+\frac{2M}{r}\right)\equiv PA_{0},\\
&h^2(r)=\frac{r^2\left(1-\frac{2M}{r}\right)}{\left(1-\frac{2M}{r}-j\right)^2}
\left(1-\frac{\omega_{p}^2(r)}{\omega_{0}^2}\left(1-\frac{2M}{r}-j\right)\right).
\end{align}

Deflection angle formula for the HT metric under these assumptions yields
\begin{equation}
\alpha_{HT0}=\pm 2\int_{R}^{\infty}f_{HT0}(r)dr - \pi,
\end{equation}
where
\begin{gather}\label{f_HT0}
f_{HT0}(r)=\frac{1}{\sqrt{ACP_{0}}}\\
\times\left(\frac{ACP_{0}
\left(1-\frac{\omega_{p}^2(r)}{\omega_{0}^2}A_{0}\right)}{A_{0}^2\left(\frac{P(R)}{A(R)}-\frac{P(r)}{A(r)}\pm h(R)\right)^2}-1\right)^{-1/2}.\nonumber
\end{gather}

This part of the deflection angle in the HT metric in the weak field approximation in vacuum gives (see Appendix~\ref{App_HT_weak})
\begin{align}\label{alpha_HT0}
\alpha_{HT0}&=\frac{4M}{R}+\frac{M^2}{R^2}\left(\frac{15\pi}{4}-4\right) \mp\frac{4J}{R^2}\\
&\mp\frac{MJ}{R^3}(10\pi-16)
+\frac{4J^2}{MR^3}+\frac{M^3}{R^3}\left(\frac{122}{3}-\frac{15\pi}{2}\right).\nonumber
\end{align}
The formula is the same as deflection angle formula (\ref{alpha_Kerr_transf}) obtained for the Kerr metric after the coordinate transformation (\ref{transf_Kerr}) and considering that $J=-Ma$.

To see the effect of the quadrupole moment in the lowest order, after neglecting the mixed terms proportional to both $\frac{J}{MR}$ and $\frac{Q}{R^3}$, i.e., $\propto\frac{JQ}{MR^4}$, we further add
\begin{alignat}{2}
&A(r) =A_{0}-KQ_{2}^2,\\
& A(r)B(r) = AB_{0},\\
&A(r)C(r)+P^2(r)=ACP_{0}-K\frac{2Mr^2}{\sqrt{r(r-2M)}}Q_{2}^1,\\
&1-\frac{\omega_{p}^2(r)}{\omega_{0}^2}A(r)=1-\frac{\omega_{p}^2(r)}{\omega_{0}^2}\left(A_{0}-KQ_{2}^2\right),\\
&\frac{P(r)}{A(r)}=PA_{0},\\
&h^2(r)=\frac{ACP_{0}-K\frac{2Mr^2}{\sqrt{r(r-2M)}}Q_{2}^1}{\left(A_{0}-KQ_{2}^2\right)^2}\\
&\times\left(1-\frac{\omega_{p}^2(r)}{\omega_{0}^2}\left(A_{0}-KQ_{2}^2\right)\right).\nonumber
\end{alignat}

Then for the deflection angle in the HT metric we obtain
\begin{equation}
\alpha_{HT}=\pm 2\int_{R}^{\infty}f_{HT}(r)dr - \pi,
\end{equation}
where
\begin{widetext}
\begin{gather}\label{f_HT}
f_{HT}(r)=\sqrt{\frac{AB_{0}}{ACP_{0}-K\frac{2Mr^2}{\sqrt{r(r-2M)}}Q_{2}^1}}
\left(\frac{\left(ACP_{0}-K\frac{2Mr^2}{\sqrt{r(r-2M)}}Q_{2}^1\right)\left(1-\frac{\omega_{p}^2(r)}{\omega_{0}^2}\left(A_{0}-KQ_{2}^2\right)\right)}{\left(A_{0}-KQ_{2}^2\right)^2\left(PA_{0}(R)-PA_{0}(r)\pm h(R)\right)^2}-1\right)^{-1/2}.
\end{gather}
\end{widetext}
Notice that $Q_{2}^1$ and $Q_{2}^2$ explicitly appearing in the last formula are functions of $r$. The function $h(R)$ is also the function of $Q_{2}^1$ and $Q_{2}^2$, but these are now functions of $R$. The lowest term with the quadrupole moment in $f_{HT}(r)$ reads
\begin{gather}
f_{K}(r)=K\frac{M}{\sqrt{r(r-2M)}}Q_{2}^1(r)+K\frac{r^2}{r^2-R^2}\label{f_K}\\
\times\left(\frac{M}{\sqrt{r(r-2M)}}Q_{2}^1(r)-\frac{M}{\sqrt{R(R-2M)}}Q_{2}^1(R)\right.\nonumber\\
\left.+Q_{2}^2(R)-Q_{2}^2(r)\right)\nonumber\\
\approx
\frac{2}{5}K\left[\frac{M^4(R^4-r^2R^2-r^4)}{r^4R^4}+\frac{4M^3(r^3-R^3)}{rR^3(r^2-R^2)}\right].\nonumber
\end{gather}

Let us introduce a mixed term containing both the plasma ($\sim\omega_p$) and quadrupole ($\sim K$) parts
\begin{align}\label{f_pl}
&f_{pl}(r)=\frac{r^2}{r^2-R^2}\\
&\times\left[\frac{\omega_{p}^2(r)}{2\omega_{0}^2}\left(1-KQ_{2}^2(r)\right)-\frac{\omega_{p}^2(R)}{2\omega_{0}^2}\left(1-KQ_{2}^2(R)\right)\right]\nonumber\\
\approx&\frac{r^2(\omega_{p}^2(r)-\omega_{p}^2(R))}{2\omega_{0}^2(r^2-R^2)}-\frac{4KM^3\left(R^3\omega_{p}^2(r)-r^3\omega_{p}^2(R)\right)}{5\omega_{0}^2rR^3(r^2-R^2)}\nonumber\\
=&f_{pKerr}(r)-f_{pHT}(r).\nonumber
\end{align}

The obtained expressions were calculated under the simplifications which arise from relations
\begin{align}
  &\frac{2M}{\sqrt{r(r-2M)}}Q_{2}^1-Q_{2}^2=\frac{3(2M^2-r^2)}{2M^2}\ln\left(\frac{r}{r-2M}\right) \\
  &-\frac{2M^2-3rM-3r^2}{rM},\nonumber\\
 & \frac{2M}{\sqrt{r(r-2M)}}Q_{2}^1=\frac{3(M-r)}{M}\ln\left(\frac{r}{r-2M}\right)\\
  &+\frac{2M^2-12rM+6r^2}{r(r-2M)}, \nonumber\\
  &Q_{2}^2=\frac{3r(r-2M)}{2M^2}\ln\left(\frac{r}{r-2M}\right)\\
  &+\frac{(r-M)(2M^2+6rM-3r^2)}{rM(r-2M)},\nonumber
\end{align}
and in the weak field approximation ($M/r\ll1$) one gets
\begin{align}
  \frac{2M}{\sqrt{r(r-2M)}}Q_{2}^1&\approx\frac{4}{5}\frac{M^4}{r^4}, \quad Q_{2}^2\approx\frac{8}{5}\frac{M^3}{r^3}.
\end{align}
When the partial deflection angle formula for the HT metric (\ref{alpha_HT0}) obtained above is taken into account in the weak field approximation, the corresponding deflection angle formula in the HT metric in plasma can be expressed as
\begin{align}
\alpha_{HT}&=\pm 2\int_{R}^{\infty}f_{HT}(r)dr - \pi =\pm 2\int_{R}^{\infty}\left[f_{HT0}(r)\right.\\
&\left.+\frac{R}{r\sqrt{r^2-R^2}}\left(f_{K}(r)+f_{pl}(r)\right)\right]dr- \pi,\nonumber
\end{align}
where individual terms are given by (\ref{f_HT0}), (\ref{f_K}), and (\ref{f_pl}).

The additional terms give
\begin{align}
&\int_{R}^{\infty}\frac{R}{r\sqrt{r^2-R^2}}f_{K}(r)dr\\
&=\frac{2}{5}K\int_{R}^{\infty}\left[\frac{M^4(R^4-r^2R^2-r^4)}{r^5R^3\sqrt{r^2-R^2}}+\frac{4M^3(r^3-R^3)}{r^2R^2(r^2-R^2)^{3/2}}\right]dr\nonumber\\
&=\frac{2}{5}K\left(-\frac{9\pi M^4}{16R^4}+\frac{8M^3}{R^3}\right),\nonumber\\
&\int_{R}^{\infty}\frac{R}{r\sqrt{r^2-R^2}}f_{pKerr}(r)dr\\
&=\frac{R}{2\omega_{0}^2}\int_{R}^{\infty}\frac{r(\omega_{p}^2(r)-\omega_{p}^2(R))dr}{(r^2-R^2)^{3/2}}=\frac{1}{2}\alpha_{refr}(R),\nonumber\\
&\int_{R}^{\infty}\frac{R}{r\sqrt{r^2-R^2}}f_{pHT}(r)dr\\
&=\frac{4KM^3}{5\omega_{0}^2}\int_{R}^{\infty}\frac{\left(R^3\omega_{p}^2(r)-r^3\omega_{p}^2(R)\right)dr}{r^2R^2(r^2-R^2)^{3/2}}\nonumber\\
 &=\frac{1}{2}\alpha_{refrHT}(R).\nonumber
\end{align}
Notice that when plasma is homogeneous, i.e., $\omega_p(r)=\omega_p(R)$, $\alpha_{refr}$ vanishes.

Complete form of the deflection angle formula in the HT metric is thus
\begin{equation}\label{alpha_HT_complete}
  \alpha=\alpha_{HT0}+\frac{32}{5}\frac{KM^3}{R^3}-\frac{9\pi}{20}\frac{KM^4}{R^4}+\alpha_{refr}(R)-\alpha_{refrHT}(R).
\end{equation}

To give a specific example, assume the plasma frequency of the form $\omega_{p}^2(r)=\mathcal{C}r^{-k}$, where $\mathcal{C}$ and $k$ are constants. Let us assume that the last two terms of (\ref{alpha_HT_complete}) can be expressed by (\ref{alpha_refr_b}) and (\ref{alpha_refrHT_dr}), respectively. The corresponding deflection angle terms then are
\begin{align}
\alpha_{refr}(R)&=-\frac{R\mathcal{C}k}{\omega_{0}^2}\int_{R}^{\infty}\frac{r^{-k-1}dr}{(r^2-R^2)^{1/2}}\label{gamma1}\\
&=\frac{\mathcal{C}k}{\omega_{0}^2R^{k}}\int_{0}^{1}\frac{u^kdu}{(1-u^2)^{1/2}}=\frac{\mathcal{C}\sqrt{\pi}}{\omega_{0}^2R^{k}}\frac{\Gamma\left(\frac{k}{2}+\frac{1}{2}\right)}{\Gamma\left(\frac{k}{2}\right)},\nonumber\\
\alpha_{refrHT}(R)&=-\frac{8KM^3R\mathcal{C}(k+3)}{5\omega_{0}^2}\int_{R}^{\infty}\frac{r^{-k-4}dr}{(r^2-R^2)^{1/2}}\label{alph_refrHT}\\
&=\frac{8KM^3\mathcal{C}(k+3)}{5\omega_{0}^2R^{k+3}}\int_{1}^{0}\frac{u^{k+3}du}{(1-u^2)^{1/2}}\nonumber\\
&=\frac{8KM^3\mathcal{C}\sqrt{\pi}}{5\omega_{0}^2R^{k+3}}\frac{\Gamma\left(\frac{k}{2}+2\right)}{\Gamma\left(\frac{k}{2}+\frac{3}{2}\right)},\nonumber
\end{align}
where
\begin{gather}
    \Gamma(z)=\int_{0}^{\infty}t^{z-1}e^{-t}dt.
\end{gather}
Expression (\ref{gamma1}) was already derived in Ref.~\cite{bisnovatyi10}. Therefore, in our approximation the deflection angle depends linearly on the quadrupole moment $Q$, as it follows from (\ref{alph_refrHT}).

\section{Deflection Angle in Some Other Spacetimes with a Quadrupole Moment}
Having obtained a general formula and an approximate result for the deflection angle in the HT metric, we shall now indicate how this procedure can be applied to other metrics with a quadrupole moment. For this reason, let us consider the Erez-Rosen (ER) metric. It reads (e.g.,
Ref.~\cite{quevedo89})
\begin{gather}
  ds^2=-e^{2\psi}dt^2+e^{2(\gamma-\psi)}\left[\left(1+\frac{M^2\sin^2\vartheta}{r^2-2Mr}\right)dr^2\right.\\
  \left.+(r^2-2Mr+M^2\sin^2\vartheta)d\vartheta^2\right] \nonumber\\
  +e^{-2\psi}(r^2-2Mr)\sin^2\vartheta d\varphi^2,\nonumber
\end{gather}
where
\begin{widetext}
\begin{align}
  \psi&=\frac{1}{2}\ln\left(1-\frac{2M}{r}\right)+\frac{q_{ER}}{2}P_2(\cos\vartheta)\left[\left(\frac{3r^2}{2M^2}-\frac{3r}{M}+1\right)\ln\left(1-\frac{2M}{r}\right)+\frac{3r}{M}-3\right],\\
  \gamma&=\frac{1}{2}\ln\left(\frac{r^2-2Mr}{r^2-2Mr+M^2\sin^2\vartheta}\right)+q_{ER}\left[\ln\left(\frac{r^2-2Mr}{r^2-2Mr+M^2\sin^2\vartheta}\right)-\frac{3}{2}\left(\frac{r}{M}-1\right)\ln\left(1-\frac{2M}{r}\right)-3\right].
\end{align}
\end{widetext}
The terms of the second order in quadrupole parameter $q_{ER}$ were omitted. Equation (\ref{ohyb_uhel}) can be used for the static metric when setting $P(r)=0$. A complete formula of the deflection angle in the ER metric in the equatorial plane thus has the form
\begin{widetext}
\begin{align}
  \alpha=&\int_{R}^{\infty}\frac{e^{\gamma}\sqrt{r^2-2Mr+M^2}}{r^2-2Mr}\left(\frac{e^{-2\psi(r)}(r^2-2Mr)\left(e^{-2\psi(r)}-\frac{\omega^2_{p}(r)}{\omega^2_0}\right)}{e^{-2\psi(R)}(R^2-2MR)\left(e^{-2\psi(R)}-\frac{\omega^2_{p}(R)}{\omega^2_0}\right)}-1\right)^{-1/2}dr-\pi.
\end{align}
\end{widetext}

To see how this formula is related to the deflection angle in the HT metric, let us express the relevant terms of the ER metric in the weak field approximation. Assuming $M/r\ll1$, these terms simplify to
\begin{align}
  A(r)&=1-\frac{2M}{r}+\frac{2q_{ER}}{15}\frac{M^3}{r^3}-\frac{4q_{ER}}{15}\frac{M^4}{r^4},\\
  B(r)&=1+\frac{2M}{r}+\frac{4M^2}{r^2}-\frac{2q_{ER}}{15}\frac{M^3}{r^3}-\frac{4q_{ER}}{15}\frac{M^4}{r^4},\\
  C(r)&=r^2\left(1-\frac{2q_{ER}}{15}\frac{M^3}{r^3}\right).
\end{align}
Setting (see, e.g., Ref.~\cite{boshkayev20ER})
\begin{equation}
  q_{ER}=-\frac{15}{2}\frac{Q}{M^3},
\end{equation}
we get the same form of terms as are the HT metric terms in the weak field approximation when $J=0$ and $Q$ is identified as a quadrupole moment (introduced in~(\ref{HT_metrika})). It can thus be seen that if in the formula for the deflection angle in the HT metric (\ref{alpha_HT_complete}) one sets $J=0$, one gets the relation for the ER metric.\\

Another exact solution of the Einstein equations with a quadrupole moment is given by the so-called $q$-metric (e.g., Ref.~\cite{toktarbay14}). It can be written as
\begin{widetext}
\begin{align}
  ds^2=&-\left(1-\frac{2\mathcal{M}_q}{r}\right)^{q+1}dt^2+\left(1-\frac{2\mathcal{M}_q}{r}\right)^{-q}
  \left[\left(1+\frac{\mathcal{M}_q^2\sin^2\vartheta}{r^2-2\mathcal{M}_qr}\right)^{-q(2+q)}\left(\frac{dr^2}{1-\frac{2\mathcal{M}_q}{r}}+r^2d\vartheta^2\right)+r^2\sin^2\vartheta d\varphi^2\right].
\end{align}
\end{widetext}

Since the other metrics are determined in terms of the 0th and 2nd multipole moments, we write down relations between them and the parameters $\mathcal{M}_q$ and $q$ (cf Ref.~\cite{frutos18}) as
\begin{gather}
M=(1+q)\mathcal{M}_q, \quad Q=-\frac{\mathcal{M}_q^3}{3}q(1+q)(2+q).
\end{gather}
These allow us to compare results obtained for the $q$-metric with those derived in other spacetimes. For this reason, it is further useful to express the $q$-metric parameters as functions of multipole moments, which read
\begin{gather}
\mathcal{M}_q=\sqrt{3\frac{Q}{M^3}+1}, \quad q=\frac{1}{\sqrt{3\frac{Q}{M^3}+1}}-1.
\end{gather}
Such defined relations set restriction on possible values of $M$ and $Q$ to guarantee that the root in both expressions is positive. For more details about the transformation between the two sets of parameters, see Ref.~\cite{memmen21}.

In the equatorial plane, the deflection angle formula reads
\begin{widetext}
\begin{gather}
  \alpha=\int_{R}^{\infty}\frac{1}{r}\left(1-\frac{2\mathcal{M}_q}{r}\right)^{-1/2}\left(1+\frac{\mathcal{M}_q^2}{r^2-2\mathcal{M}_qr}\right)^{-\frac{q(2+q)}{2}}
 \left(\frac{r^2\left(1-\frac{2\mathcal{M}_q}{r}\right)^{-q}\left(\left(1-\frac{2\mathcal{M}_q}{r}\right)^{-q-1}-\frac{\omega^2_{p}(r)}{\omega^2_0}\right)}{R^2\left(1-\frac{2\mathcal{M}_q}{R}\right)^{-q}\left(\left(1-\frac{2\mathcal{M}_q}{R}\right)^{-q-1}-\frac{\omega^2_{p}(R)}{\omega^2_0}\right)}-1\right)^{-1/2}dr-\pi.
\end{gather}
\end{widetext}
For $q=0$ the formula is identical with that for the Schwarzschild metric, see Ref.~\cite{perlick15}. Hence, obtaining the form of the deflection angle in the weak field approximation for the HT metric yields the deflection angle in the $q$-metric as in the case of the ER metric.

\section{Illustration of the Deflection Angles}
As was shown above, there is an evident connection between the Kerr, Hartle-Thorne, Erez-Rosen, and $q$-metric. They can all be reduced to the Schwarzschild metric if certain parameters are neglected. When comparing the deflection angles in plasma in these spacetimes, results obtained for the HT metric can be applied for the other cases under certain simplifications. In this section, the results presented above mainly analytically are illustrated graphically.

It is evident that the applications of the deflection angle formulae in the weak field approximation are limited and cannot be used too close to a strongly gravitating object. It is thus desirable to compare the results with exact formulae and set a range of suitable radial distances where the weak field approximation is safe to use.

To quantitatively express how much the plasma profiles manifest, several were chosen and the results were depicted along with the deflection angle in vacuum.

Further, we compare the ray trajectories in vacuum and in various plasma models. It can thus be directly seen how much different parameters of a gravitating source affect the ray propagation.

Since the quadrupole moment substantially influences the results, solutions for several explicit quadrupole moments are considered and discussed.

\subsection{Comparison of approximate formulae and exact results}
Let us first show how the weak field approximation formulae differ from the exact results in vacuum. Since plasma presence causes a decrease of the deflection angles, the vacuum case represents an upper estimate. This provides an idea when it is indeed necessary to apply an exact formula. For sufficiently large $r\gg M$, terms proportional to other parameters but $M$ become negligible and all solutions basically take form valid for the Schwarzschild metric. So, let us consider the Schwarzschild metric first.

\begin{figure}
  \centering
  \includegraphics[width=0.995\columnwidth]{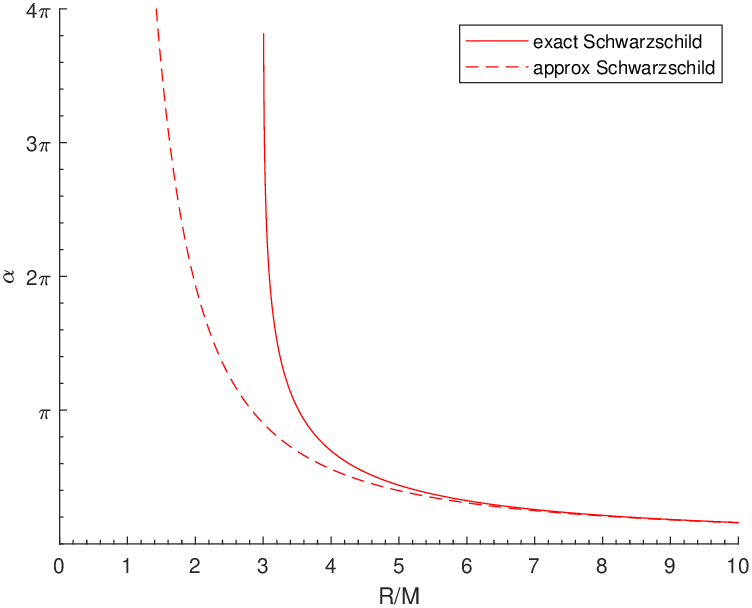}
  \caption{Exact and approximate deflection angles in the Schwarzschild metric as a function of ray minimal radial distance $R$. The solid curve shows the deflection angle calculated by the exact general formula (\ref{ohyb_uhel}) applied under assumption $P(r)=0$, while the approximate solution derived in the weak field approximation is depicted by the dashed curve.}\label{Schw_pribl}
\end{figure}

Fig.~\ref{Schw_pribl} shows the dependence of the exact and approximate deflection angles in the Schwarzschild metric in vacuum as functions of the ray closest radial distance $R$. The exact deflection angles, obtained by using formula~(\ref{ohyb_uhel}) with $P(r)=P(R)=0$, are drawn by the solid curve, whereas the approximate expression calculated from (\ref{alpha_BL}) under assumption $a=0$ is shown by the dashed curve. Fig.~\ref{Schw_pribl} demonstrates that at sufficiently large radial distances (from at about $6M$), the approximate deflection angle formula can be applied with a satisfactory accuracy. Moreover, it is also nicely seen that while the exact solution at small radial distances asymptotically reaches $3M$ (photon sphere radius), the approximate solution diverges for small $R$.

\begin{figure*}
  \centering
  \includegraphics[width=1\textwidth]{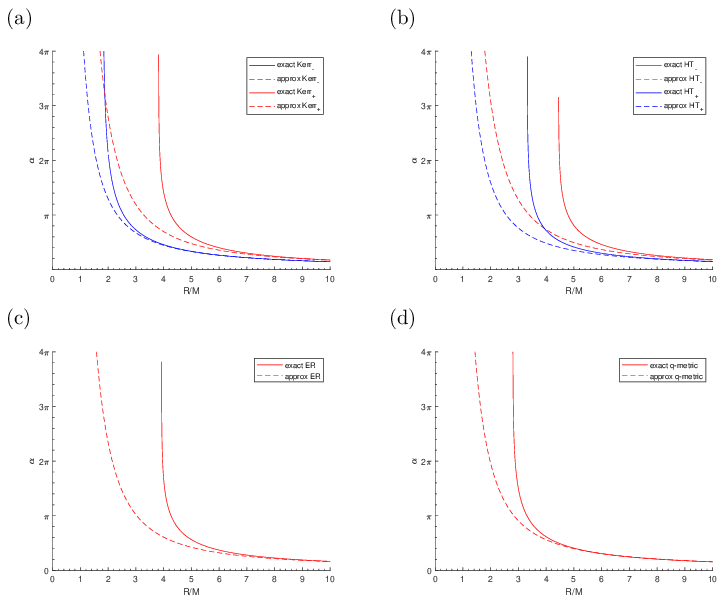}
  \caption{Exact and approximate deflection angles for the (a) Kerr metric with $J=0.8$, (b) HT metric with $J=0.8$, $Q=2.5$, (c) ER metric with $q_{ER}=-18.75$, and (d) $q$-metric with $q=0.25$.}\label{def_angle_pribl}
\end{figure*}

Fig.~\ref{def_angle_pribl} shows a comparison of the exact and approximate deflection angles in the Kerr metric (panel \ref{def_angle_pribl}a), the HT metric (panel \ref{def_angle_pribl}b), the ER metric (panel \ref{def_angle_pribl}c), and the $q$-metric (panel \ref{def_angle_pribl}d). The dashed and solid curves have the same meaning as in Fig.~\ref{Schw_pribl}. Because of the presence of the angular momentum, the Kerr and HT metrics exhibit two branches of solutions, corresponding to the co-rotating (+)/counter-rotating (-) orbits, respectively.

In Fig.~\ref{def_angle_pribl}a it can be seen that in the Kerr metric, the sufficient correspondence between the exact and approximate solutions occurs already at around $4M$ in branch with a negative sign of $a$, which is significantly closer to the photon sphere radius than in both the Schwarzschild and positive branch cases. On the contrary, solution with a negative sign gives worse accuracy of the approximate solution in the HT metric (see Fig.~\ref{def_angle_pribl}b). This is caused by different sign in the definition of $J$ in the Kerr and HT metric (for the correspondence between the Kerr and HT metric one has to set $J=-Ma$, see Ref.~\cite{HartleThorne1968}). Hence, the branches are transposed. The angular momentum values were set to be equal to $J=0.8$ in both cases to directly expose the effect of the quadrupole moment. The quadrupole moment in Fig.~\ref{def_angle_pribl}b was chosen to be $Q=2.5$. In comparison with the Kerr metric, due to a presence of the quadrupole moment the position of the photon sphere radius is further in radial distances for both branches in the HT metric case. In the Kerr metric the correspondence between the approximate and exact solutions occurs already around $4M$ in the negative branch and around $7M$ in the positive branch. On the contrary, in the HT metric the coincidence of both solutions in the negative branch is at around $8M$ and in the positive branch it is at around $6M$.

Comparisons between the exact and approximate vacuum results in the ER metric and $q$-metric are shown in Fig.~\ref{def_angle_pribl}c and \ref{def_angle_pribl}d, respectively.  The quadrupole coefficients in both cases can be easily directly compared with the quadrupole moment in the HT metric. Additionally, $M_q$ present in the $q$-metric was calculated from the defined $M$ which thus remains the same for all discussed metrics ($M=1$). It is seen that even the corresponding quadrupole moments can have different effects in particular metrics. In the ER metric the approximate and exact solutions are in a good agreement already at around $7M$, which is comparable with both branches in the HT metric. On the other hand, in the $q$-metric the sufficient correspondence between exact and approximate solutions occurs already at around $5M$. In fact, in this case the two solutions are in a good agreement and coincide much better than in the previous case. This demonstrates that not only the quadrupole moment value itself plays a significant role, but also the way how it enters into the metric is substantial. Hence, it can be seen that in the $q$-metric the quadrupole moment presence influences the spacetime in a different way than in the ER and HT metrics.

\subsection{Comparison of the deflection angles in vacuum and in plasma}
If plasma is present, the deflection angle changes, depending on the plasma properties. Following Ref.~\cite{PerlickTsupkoNN}, we assume the plasma frequency to be $\omega_p(r)^2=10\omega_0^2\left(\frac{M}{r}\right)^k$, where $k=3/2,5/2,7/2$, respectively, to have some specific model for illustration. Note that the results for the Schwarzschild and Kerr metrics presented here were obtained already in Ref.~\cite{PerlickTsupkoNN}. The parameter $M$ in the definition of $\omega_p(r)$ represents a mass of a given gravitating object, which appears in the expansion terms in the weak field approximation. (This is the 0th multipole moment as given in the Geroch's definition, cf Ref.~\cite{geroch70}.)

\begin{figure}
  \centering
  \includegraphics[width=1\columnwidth]{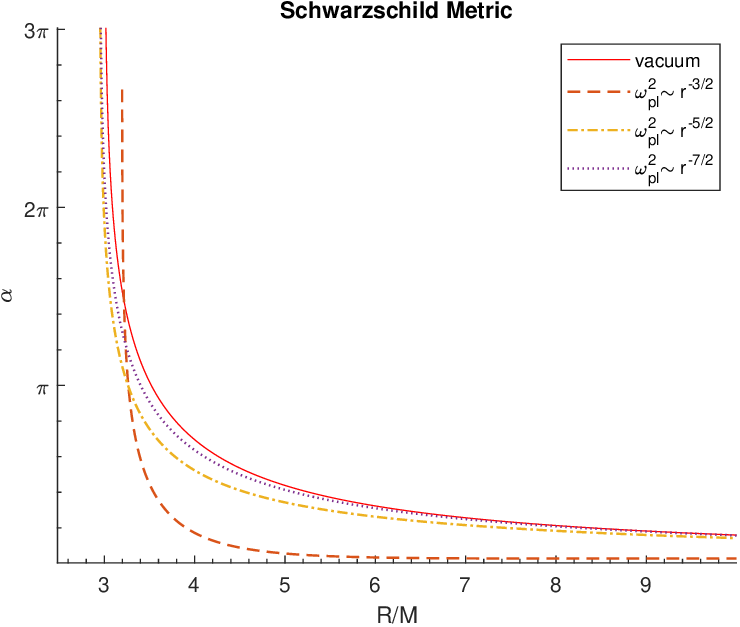}
  \caption{Deflection angle for the Schwarzschild metric in vacuum (red curve) and in several plasma cases (dashed curves).} \label{Schw_pl_vs_vac}
\end{figure}

\begin{figure*}
  \centering
  \includegraphics[width=1\textwidth]{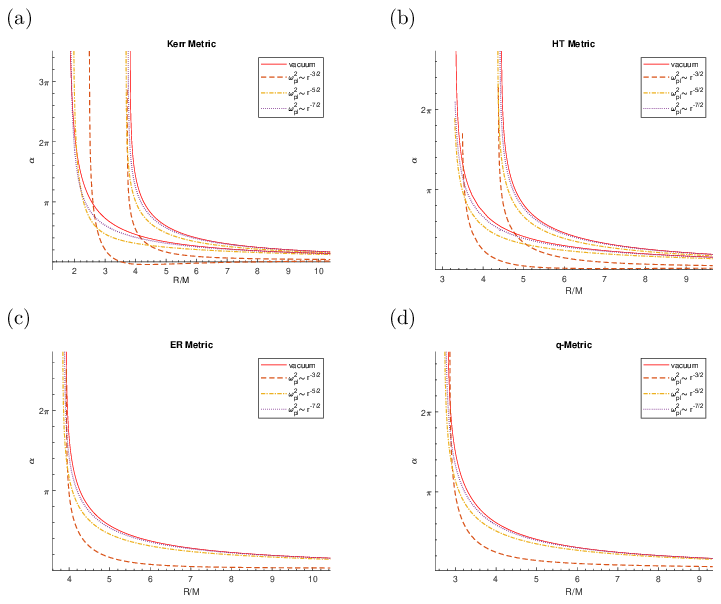}
  \caption{Deflection angle in vacuum and plasma cases for the (a) Kerr metric with $J=0.8$, (b) HT metric with $J=0.8$, $Q=2.5$, (c) ER metric with $q_{ER}=-18.75$, and (d) $q$-metric with $q=0.25$. In panel (a), the branch closer to $R/M=0$ corresponds to the rays with negative angular momentum, while the other is with the positive angular momentum. In panel (b), the rays with negative angular momentum occur in the branch further from the origin.}\label{def_angle_pl_vs_vac}
\end{figure*}

The deflection angle $\alpha$ as a function of $R$ in the Schwarzschild metric in vacuum and in plasma with different values of coefficient $k$ is shown in Fig.~\ref{Schw_pl_vs_vac}. Exact vacuum solution is drawn by the red solid curve, while the deflection angles when plasma is present are drawn by the dashed curves. As already mentioned above, Fig.~\ref{Schw_pl_vs_vac} (and all panels in Fig.~\ref{def_angle_pl_vs_vac}) demonstrates that the plasma presence causes the light rays to be less bent than in vacuum. This is due to the fact that the refractive index in plasma is $<1$ (in other non-vacuum media with $n>1$ the rays would be more bent). It is also seen that when the plasma frequency (or rather the plasma density\footnote{Considering that it holds $\omega_p(r)^2=\mathcal{C}_eN(r)$, where $\mathcal{C}_e$ is a constant factor and $N(r)$ usually stands for electron density.}) decreases steeper with a radial distance, its effect is less manifested than for a gradual decrease. Because for chosen plasma frequency functions at larger $r$ the plasma frequency sufficiently decreases, the deflection angles in plasma become close to deflection angles in vacuum. At small radial distances, the deflection angles are limited by the photon sphere radius. In the Schwarzschild metric the range of radial distances where the plasma effect manifests most is thus between around $3M$ and $5M$, depending on a concrete plasma profile.

In plasma with $k=3/2$ profile, the deflection angles steeply decrease already at low $R$, close to the photon sphere radius ($3M$), and they are significantly lower than in the vacuum or other plasma cases. Plasma presence can thus significantly influence the deflection angles at substantially large radial distances and this effect can also persist for closest distances. However, the plasma frequency must evolve reasonably. Hence, even the rays which are eventually less influenced by gravity (their closest radial distance is large) can still be significantly affected by plasma.

Similar results for the Kerr and HT metrics are shown in Fig.~\ref{def_angle_pl_vs_vac}a,b. The effect of plasma frequency with $k=3/2$ is again substantially stronger than in the other cases. While the presence of a quadrupole moment in the HT metric causes a quantitative change of the deflection angles (they are in general larger than in the Kerr metric), there is no significant qualitative change. Range of the deflection angles where the effect of plasma is manifested most, extends from the photon sphere radius to several closest radial distances with corresponding values in both solution branches. While the plasma effect in the solution with the negative angular momentum in the Kerr metric (analogous to positive angular momentum in the HT metric) is most obvious at radial distances up to around $6M$, in the Kerr solution with the positive angular momentum (corresponding to negative in the HT metric) it is spread slightly further, approximately between $7M$ and $8M$.

Notice that in the negative branch of the Kerr metric solution with plasma with $k=3/2$ profile, the deflection angles are negative at small $R$. This effect is apparent up to radial distances around $6M$. This means that light is actually bent to the opposite side and plasma thus causes a significant effect. Although for the simple plasma profiles, the negative deflection angles are small, this effect could be more apparent in some other types of plasmas.

Deflection angles in vacuum and various plasma cases for the ER metric and $q$-metric are shown in Fig.~\ref{def_angle_pl_vs_vac}c,d, respectively. While the deflection angle profiles obtained for the ER metric slightly vary from the Schwarzschild  case, the $q$-metric exhibits almost identical behavior. The difference between the deflection angles in plasma and in vacuum is less significant and the effect of plasma with $k=3/2$ profile is even not as peculiar as in the other cases. Fig.~\ref{def_angle_pl_vs_vac}d thus indisputably demonstrates the robustness of the $q$-metric definition, where the quadrupole moment plays the most significant role. The effect of plasma is significantly less manifested in the $q$-metric than in other metrics. However, it still holds that in plasma with $k=3/2$ profile the deflection angles are noticeably lower at substantially large radial distances.

\subsection{Ray trajectories in vacuum and in plasma}
How specific rays around given gravitating objects look like is shown below. The same plasma profiles as in the previous section are used, only the case with $k=3/2$ is not considered because for $k=3/2$ the plots are rather messy.

The equations of motion (\ref{eom1}),(\ref{eom2}) and corresponding equations for $p_\varphi$ and $p_r$ were used to calculate the individual ray trajectories. The rays start at the same initial point and with four different impact parameters (defined by Eq.~(\ref{def_b})); their values are $2,5,8,10$. The connection between impact parameter $b$ and the closest radial distance $R$ used in the previous sections is discussed in detail in Appendix~\ref{App_b}.

\begin{figure}
  \centering
  \includegraphics[width=1\columnwidth]{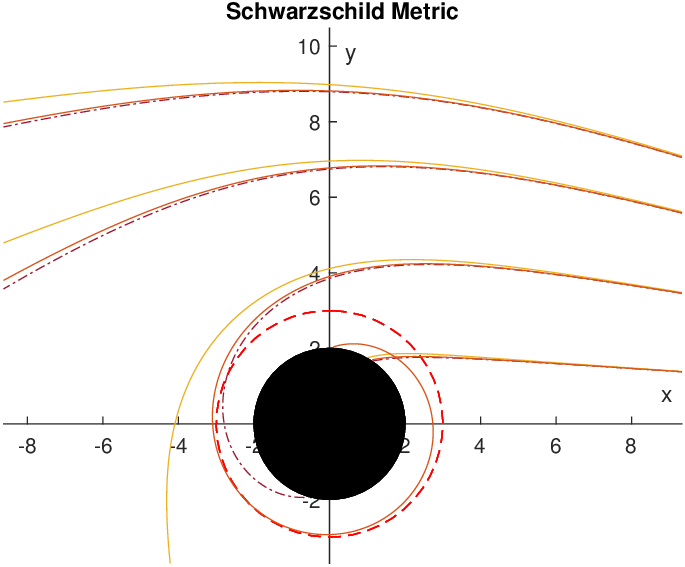}
  \caption{Ray trajectories in the vicinity of the Schwarzschild black hole in vacuum (dash-dotted lines) and plasma cases (solid lines) when $k=5/2$ (orange) and $k=7/2$ (yellow). The dashed circle shows position of the radius of the circular photon orbit. Ray impact parameter were set as $2,5,8,10$.}\label{Schw_traject}
\end{figure}

\begin{figure*}
  \centering
  \includegraphics[width=1\textwidth]{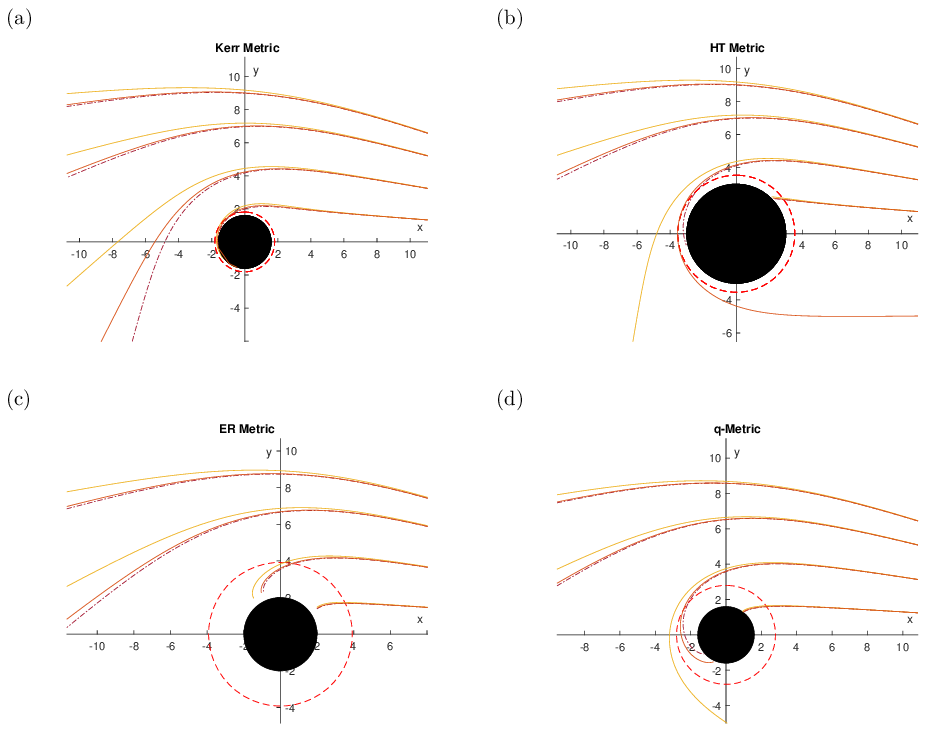}
  \caption{Same as Fig.~\ref{Schw_traject}, but for the (a) Kerr metric with $J=0.8$, (b) HT metric with $J=0.8$, $Q=2.5$, (c) ER metric with $q_{ER}=-18.75$, and (d) $q$-metric with $q=0.25$.}\label{rest_traject}
\end{figure*}

Ray trajectories around the discussed metrics are presented in Figs.~\ref{Schw_traject},\ref{rest_traject}, respectively. Rays in vacuum are drawn by dash-dotted purple curves, while those in plasma are plotted by the solid curves. Trajectories obtained in plasma with a density profile with $k=5/2$ are orange and those with $k=7/2$ are yellow curves. Since $n<1$, the rays in plasma are generally less bent than in the vacuum case. Moreover, the higher coefficient $k$ is, the less the rays are bent. This is well demonstrated in all cases. Red dashed circles in Figs.~\ref{Schw_traject},\ref{rest_traject} show the position of the circular photon orbits. Due to the rotation, the Kerr and HT metrics (Figs.~\ref{rest_traject}a,b, respectively) have actually two circular photon orbits. Only photon orbits with a positive angular momentum (in terms of the HT metric) are plotted. Sizes of gravitating objects in Figs.~\ref{Schw_traject},\ref{rest_traject} (radii of black circles) are defined as their event horizons.

Fig.~\ref{Schw_traject} shows the rays in the Schwarzschild metric. It can be seen that rays with the same impact parameters will or will not be captured by the black hole, depending on the plasma profile. This is also manifested in the case of the HT metric in Fig.~\ref{rest_traject}b. For sufficiently small impact parameters the plasma presence cannot prevent the rays from being captured by gravitating object.

The ray trajectories in the other metrics are shown in Fig.~\ref{rest_traject}. Generally, it can be seen that the behaviour of the rays significantly varies and it is well demonstrated that plasma presence can play a significant role. It is also apparent that along with the increasing impact parameters the effect of plasma decreases, since both the gravitation and plasma density are weaker and they hence less influence the light trajectories.

Figs.~\ref{rest_traject}a,b are useful to demonstrate the effect of the quadrupole moment. While maintaining the same angular moment, it is evidently seen that the object described by the HT metric is not only significantly larger than the Kerr metric, but it also significantly more affects the ray trajectories with larger impact parameters. Sufficiently large quadrupole moment thus can be strongly manifested also at larger radial distances.

Quadrupole moment effects are further well demonstrated in comparison with Figs.~\ref{rest_traject}c,d, which show the ray trajectories in the vicinity of the ER metric and $q$-metric, respectively. It is clearly seen that the quadrupole moment presence again causes the photon sphere radius to be larger and the more distant rays are thus significantly affected.

The parameters of the gravitating objects are maintained the same as in previous sections. However, in the $q$-metric with $q=2.5$ all investigated rays fall on the object and the results hence are not particularly interesting. The value of the quadrupole parameter was set to be $0.85$ in this case (Fig.~\ref{rest_traject}d).

\subsection{Effect of the quadrupole moment on the light rays}
\begin{figure*}
  \centering
  \includegraphics[width=1\textwidth]{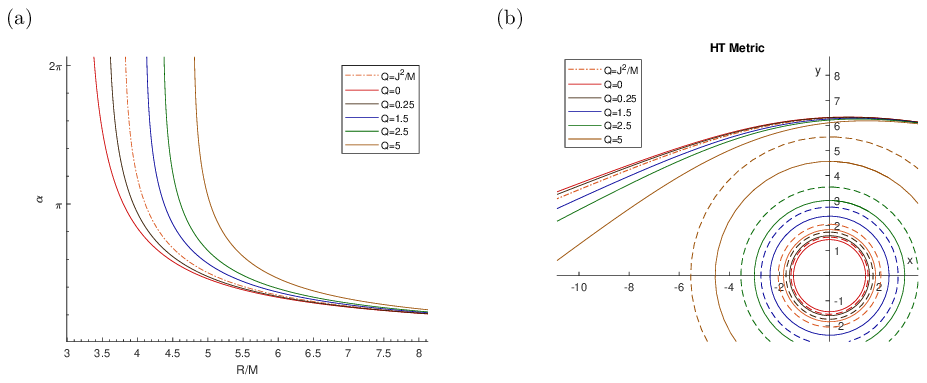}
  \caption{Effect of the quadrupole moment in the HT metric with $J=0.8$ on the (a) deflection angle, (b) ray trajectories in plasma. Plasma defined by $\omega_p(r)$ with $k=5/2$ was assumed. Ray impact parameter is equal to 7.}\label{ruzna_q}
\end{figure*}

To better demonstrate how the rays evolve in the HT metric, the astrophysically most relevant metric with a quadrupole moment, we discuss its effect in more detail. Different quadrupole moments will have different impact on the light propagation. This is demonstrated in Figs.~\ref{ruzna_q}a,b. Both the deflection angles corresponding to different quadrupole moments (Fig.~\ref{ruzna_q}a) and the ray trajectories around the HT metric with various quadrupole moments (Fig.~\ref{ruzna_q}b) are shown. Only results with the negative angular momentum are plotted in Fig.~\ref{ruzna_q}, and only plasma of density profile with $k=5/2$ was considered.

The deflection angle in the Kerr metric, when $Q=J^2/M$, is shown by the dash-dotted curve. Notice that in the comparison with the other quadrupole moments, the Kerr metric case plays no special role. It simply represents one concrete choice of the quadrupole moment. Moreover, as it can also be seen in  Fig.~\ref{ruzna_q}b, the larger quadrupole moment causes the photon sphere radius to increase and the deflection angle at larger $R$ hence also increases.

Fig.~\ref{ruzna_q}b shows how specific rays evolve around an object with different quadrupole moments described by the HT metric in plasma. Because both the event horizon and the photon sphere radius differ for various quadrupole moments, their radii are plotted by different colors as solid and dashed circles, respectively. In this case, the ray impact parameter was set to 7.

\section{Results and discussion}
Besides mass and angular momentum, one of the basic features which characterize a gravitating object is the quadrupole moment. The effect of this parameter on the light propagation around such an object with surrounding plasma was investigated in the present work.

We first derived the general formula describing the deflection angle in stationary axially symmetric spacetimes with a refractive and dispersive medium in a cold plasma approximation in the equatorial plane. This formula can also be used for a spherically symmetric object, which leads to an expression already obtained in Ref.~\cite{perlick15}. Our formula (\ref{ohyb_uhel}) actually holds in a more general situation when an arbitrary dispersive medium around a gravitating object is present, i.e., it satisfies $n=n(r,\omega(r))$. In this sense, a cold non-magnetized plasma applied further in this study is just a special example of such environment (cf Ref.~\cite{tsupko21}).

The expression for the deflection angle was applied in the Kerr and HT metrics to demonstrate how the presence of a quadrupole moment affects the light propagation. The results derived first in the weak field approximation, when $M/R\ll 1$, were then compared with exact expressions given by formula (\ref{ohyb_uhel}). It was demonstrated that the quadrupole moment enters the formula not until the order of $\mathcal{O}(M^3/R^3)$. Hence, the quadrupole moment manifests only at sufficiently small radial distances. The results were also presented graphically. For completeness, we graphically demonstrated all results also for an object described by the Schwarzschild metric.

To understand the effect of a quadrupole moment in other cases, the deflection angles in the Erez-Rosen metric and $q$-metric were calculated. In the $q$-metric the weak field approximation gives very good results, as it can be used at comparatively close radial distances in comparison with the other cases. However, its effect is naturally fundamentally affected by the chosen value of the quadrupole parameter which was relatively low. These examples also showed the application of the general expression for the deflection angle in spacetimes without an angular momentum.

Our results confirm that due to the medium refractive index in the cold plasma approximation, its presence effectively diminishes the light bending. Hence, rays which would be captured by a gravitating object in vacuum can escape from its vicinity in plasma.

The effect of the quadrupole moment on the light propagation in the vicinity of a gravitating object in plasma was analyzed separately. The obtained results showed that a concrete definition of the quadrupole moment in the HT metric is crucial for its corresponding manifestation in the light propagation. The comparison of the results with the Kerr metric (in which the quadrupole moment has a special value of $J^2/M$) demonstrated that in the HT metric the quadrupole moment can both decrease and increase the deflection angle.

The ray impact parameter $b$ is a suitable ray characterization because it is a frame independent dimensionless quantity. Since our results were presented first in terms of the ray closest radial distance $R$, we further found a relation between the two. A general formula between $R$ and $b$ in an axially symmetric spacetime in plasma was also derived.

We demonstrated how the quadrupole moment can significantly change the light propagation. Our approach is sufficiently universal to be applied also for other spacetimes and plasma environments.

\begin{acknowledgments}
B.B. acknowledges the support of the Charles University Grant Agency under Contract No. 317421. B.B. and J.B. were further supported by
the Czech Grant Agency under Contract No. 21/11268S.
\end{acknowledgments}

\appendix

\section{HT metric terms entering the deflection angle formula}\label{App_HT}
Assuming that the rays move in the equatorial plane, the terms in the HT metric are substantially simplified because then $\vartheta=\pi/2$ and $P_2(\cos\vartheta)|_{\vartheta=\frac{\pi}{2}}=-\frac{1}{2}$. The terms present in the deflection angle formula (\ref{ohyb_uhel}) read
\begin{align}
A(r) &=A_1\left[1-j\left(1+\frac{M}{r}\right)-KQ_{2}^2\right]-j_{1}^2\left[1\right.\\
&\left.+j\left(1+\frac{2M}{r}\right)-K\left(\frac{2M}{\sqrt{r(r-2M)}}Q_{2}^1-Q_{2}^2\right)\right],\nonumber\\
B(r) &= A_{1}^{-1}\left[1+j\left(1-\frac{5M}{r}\right)+KQ_{2}^2\right], \\
C(r) &=r^2\,\left[1+j\left(1+\frac{2M}{r}\right)\right.\\
&\left.-K\left(\frac{2M}{\sqrt{r(r-2M)}}Q_{2}^1-Q_{2}^2\right)\right],\nonumber\\
P(r) &=-rj_1\left[1+j\left(1+\frac{2M}{r}\right)\right.\\
&\left.-K\left(\frac{2M}{\sqrt{r(r-2M)}}Q_{2}^1-Q_{2}^2\right)\right].\nonumber
\end{align}
Explicit terms arising in the deflection angle formula~(\ref{ohyb_uhel}) thus look as follows:
\begin{widetext}
\begin{alignat}{2}
&A(r)B(r) =\left\{1-j\left(1+\frac{M}{r}\right)-KQ_{2}^2-j_{1}^2A_{1}^{-1}\left[1+j\left(1+\frac{2M}{r}\right) -K\left(\frac{2M}{\sqrt{r(r-2M)}}Q_{2}^1-Q_{2}^2\right)\right]\right\}\label{HT_equa1}\\
&\times\left[1+j\left(1-\frac{5M}{r}\right)+KQ_{2}^2\right],\nonumber \\
&A(r)C(r)+P^2(r)=A_1r^2\left[1-j\left(1+\frac{M}{r}\right)-KQ_{2}^2\right]\left[1+j\left(1+\frac{2M}{r}\right)-K\left(\frac{2M}{\sqrt{r(r-2M)}}Q_{2}^1-Q_{2}^2\right)\right],\\
&1-\frac{\omega_{p}^2(r)}{\omega_{0}^2}A(r)=1-\frac{\omega_{p}^2(r)}{\omega_{0}^2}\left\{A_1\left[1-j\left(1+\frac{M}{r}\right)-KQ_{2}^2\right]-j_{1}^2\left[1+j\left(1+\frac{2M}{r}\right)-K\left(\frac{2M}{\sqrt{r(r-2M)}}Q_{2}^1-Q_{2}^2\right)\right]\right\},\\
&\frac{P(r)}{A(r)}=-rj_1\left[1+j\left(1+\frac{2M}{r}\right)-K\left(\frac{2M}{\sqrt{r(r-2M)}}Q_{2}^1-Q_{2}^2\right)\right] \label{HT_equa2}\\
&\times\Biggl\{A_1\Bigl[1-j\Bigl(1+\frac{M}{r}\Bigr)-KQ_{2}^2\Bigr]-j_{1}^2\left[1+j\left(1+\frac{2M}{r}\right)-K\left(\frac{2M}{\sqrt{r(r-2M)}}Q_{2}^1-Q_{2}^2\right)\right]\Biggr\}^{-1}.\nonumber
\end{alignat}
\end{widetext}

\section{Deflection angle terms in the Kerr metric in the weak field approximation}\label{App_Kerr}
The deflection angle formula in the Kerr metric in the equatorial plane generally reads
\begin{equation}\label{App_alpha_Kerr}
\alpha=\pm 2\int_{R}^{\infty}f_{Kerr}(r)dr -\pi,
\end{equation}
where
\begin{widetext}
\begin{align}\label{App_f_Kerr}
f_{Kerr}(r)=\frac{\sqrt{r(r-2M)}}{r^2-2Mr+ a^2} \left(\frac{r^2(R-2M)^2(r^2-2Mr+a^2)\left(1-\frac{\omega_{p}^2(r)}{\omega_{0}^2}\left(1-\frac{2M}{r}\right)\right)}{\left(2Ma(R-r)\pm (r-2M)h(R)\right)^2}-1\right)^{-1/2}.
\end{align}
\end{widetext}

Although the complete calculation of the above integral (\ref{App_alpha_Kerr}) is quite complicated, it can be significantly simplified in the weak field approximation when $M/r\ll1$. Under this assumption, the terms up to the third order are further considered. Note that term $\propto$ $\frac{a^3}{R^3}$ is not present in (\ref{App_f_Kerr}) from the definition. Moreover, the HT metric is accurate up to the second order in $a$. In the weak field approximation the individual terms in \emph{vacuum} ($\omega_{p}=0$) take the form
\begin{align}
  &\frac{\sqrt{r(r-2M)}}{r^2-2Mr+a^2}\approx\frac{1}{r}\left(1+\frac{M}{r}-\frac{a^2}{r^2}+\frac{3M^2}{2r^2}+\frac{5M^3}{2r^3}-\frac{3Ma^2}{r^3}\right), \\
  &r^2(R-2M)^2(r^2-2Mr+a^2)\approx r^4R^2\left[1-\frac{2M}{rR}(2r+R)\right.\\
  &\left.+\frac{4M^2}{rR^2}(r+2R)+\frac{a^2}{r^2}-\frac{4Ma^2}{r^2R}-\frac{8M^3}{rR^2}\right],\nonumber\\
  &\left(2Ma(R-r)\pm (r-2M)h(R)\right)^2\approx r^2R^4\\
  &\times\left[1-\frac{2M}{rR}(r+2R)+\frac{4M^2}{r^2R}(2r+R)+\frac{a^2}{R^2}-\frac{4Ma^2}{rR^2}\right.\nonumber\\
  &\left.-\frac{8M^3}{r^2R}\pm\frac{4Ma}{rR^2}(R-r)\left(1-\frac{M}{rR}(r+2R)\right)\right].\nonumber
\end{align}
This leads to
\begin{gather}
  \left(\frac{r^2(R-2M)^2(r^2-2Mr+a^2)}{\left(2Ma(R-r)\pm (r-2M)h(R)\right)^2}-1\right)^{-1/2}\approx\\
  \frac{R}{\sqrt{r^2-R^2}}
  \left(1+\frac{Mr}{R(r+R)}+\frac{M^2(4R^2+4rR+3r^2)}{2R^2(r+R)^2}\right.\nonumber\\
  \left.+\frac{a^2}{2R^2}\mp\frac{2Mar}{R^2(r+R)}\left(1+\frac{M(4R^2+3rR+2r^2)}{rR(r+R)}\right)\right.\nonumber\\
  \left.+\frac{Ma^2(4R^2+2rR+3r^2)}{2rR^3(r+R)}\right.\nonumber\\
  \left.+\frac{M^3(4R(r+R)(2R^2+2rR+3r^2)+5r^4)}{2rR^3(r+R)^3}\right),\nonumber
\end{gather}
and thus
\begin{gather}
  f_{Kerr}(r)\approx \frac{R}{r\sqrt{r^2-R^2}}\left(1+\frac{M(R^2+rR+r^2)}{rR(r+R)}\right.\\
  \left.+\frac{3M^2(R^2+rR+r^2)^2}{2r^2R^2(r+R)^2}+\frac{a^2(r^2-2R^2)}{2r^2R^2}\right.\nonumber \\
  \left.\mp \frac{2Mar}{R^2(r+R)}\left(1+\frac{M(5R^2+4rR+2r^2)}{rR(r+R)}\right)\right.\nonumber\\
  \left.+\frac{3Ma^2(r^4+R(R+r)(r^2-2R^2))}{2r^3R^3(r+R)}\right.\nonumber\\
  \left.+\frac{5M^3(R^2+rR+r^2)^3}{2r^3R^3(r+R)^3}\right).\nonumber
\end{gather}

The integration of the individual terms yields
\begin{align}
 & \int_{R}^{\infty}\frac{Rdr}{r\sqrt{r^2-R^2}}=\frac{\pi}{2}, \\
 & \int_{R}^{\infty} \frac{M(R^2+rR+r^2)dr}{r^2(r+R)\sqrt{r^2-R^2}}=\frac{2M}{R},\\
& \int_{R}^{\infty} \frac{3M^2(R^2+rR+r^2)^2dr}{2r^3R(r+R)^2\sqrt{r^2-R^2}}=\frac{M^2}{8R^2}(15\pi-16),\\
&  \int_{R}^{\infty} \frac{2Madr}{R(r+R)\sqrt{r^2-R^2}}=\frac{2Ma}{R^2},\\
&  \int_{R}^{\infty} \frac{a^2(r^2-2R^2)dr}{2r^3R\sqrt{r^2-R^2}}=0,\\
&  \int_{R}^{\infty} \frac{2M^2a(5R^2+4rR+2r^2)dr}{rR^2(r+R)^2\sqrt{r^2-R^2}}=\frac{M^2a}{R^3}(5\pi-8),\\
&  \int_{R}^{\infty} \frac{3Ma^2(r^4+R(R+r)(r^2-2R^2))dr}{2r^4R^2(r+R)\sqrt{r^2-R^2}}=\frac{Ma^2}{R^3},\\
& \int_{R}^{\infty} \frac{5M^3(R^2+rR+r^2)^3dr}{2r^4R^2(r+R)^3\sqrt{r^2-R^2}}=\frac{M^3}{R^3}\left(\frac{61}{3}-\frac{15\pi}{4}\right).
\end{align}

\section{Deflection angle terms in the HT metric in the weak field approximation}\label{App_HT_weak}
As already discussed in Section~\ref{Sec_HT_weak}, the deflection angle formula for the HT metric in the equatorial plane in \emph{vacuum} and when $K=0$ yields
\begin{equation}
\alpha_{HT0}=\pm 2\int_{R}^{\infty}f_{HT0}(r)dr - \pi,
\end{equation}
where $f_{HT0}(r)$ is given by (\ref{f_HT0}) in which $\omega_{p}=0$.

When assuming the weak field approximation, i.e., $M/r\ll1$, one gets
\begin{gather}
  \frac{1}{\sqrt{r^2\left(1-\frac{2M}{r}\right)}}\approx \frac{1}{r}\left(1+\frac{M}{r}+\frac{3M^2}{2r^2}+\frac{5M^3}{2r^3}\right),
\end{gather}
and in vacuum it further holds \footnote{Let us consider that $j=\frac{J^2}{Mr^3}$ to see the explicit dependence on $r$.}
\begin{gather}
\left(\frac{\frac{r^2\left(1-\frac{2M}{r}\right)}{\left(1-\frac{2M}{r}-\frac{J^2}{Mr^3}\right)^2}}{\left(\frac{P(R)}{A(R)}-\frac{P(r)}{A(r)}\pm h(R)\right)^2}-1\right)^{-1/2}\\
\approx
\frac{R}{\sqrt{r^2-R^2}}\left(1+\frac{Mr}{R(r+R)}+\frac{M^2(4R^2+4rR+3r^2)}{2R^2(r+R)^2}\right.\nonumber\\
\left.\mp\frac{2Jr}{R^2(r+R)}\left(1+\frac{M(4R^2+3rR+2r^2)}{rR(r+R)}\right)\right.\nonumber\\
\left.+\frac{J^2(R^2+rR+r^2)}{MrR^3(r+R)}\right.\nonumber\\
\left.+\frac{M^3(8R^2(r+R)^2+12r^2R(r+R)+5r^4)}{2rR^3(r+R)^3}\right) .\nonumber
\end{gather}
Individual terms of the deflection angle thus are
\begin{gather}
f_{HT0}(r)\approx \frac{R}{r\sqrt{r^2-R^2}}\left[1+\frac{M(R^2+rR+r^2)}{rR(r+R)}\right.\\
  \left.+\frac{3M^2(R^2+rR+r^2)^2}{2r^2R^2(r+R)^2}\mp \frac{2Jr}{R^2(r+R)}\right.\nonumber\\
  \left.\times\left(1+\frac{M(5R^2+4rR+2r^2)}{rR(r+R)}\right)\right.\nonumber\\
  \left.+\frac{J^2(R^2+Rr+r^2)}{MrR^3(r+R)}+\frac{5M^3(R^2+rR+r^2)^3}{2r^3R^3(r+R)^3}\right].\nonumber
\end{gather}
It can be seen that most of the terms are the same as these obtained for the Kerr metric.
Integration of the corresponding terms gives
\begin{align}
  &\int_{R}^{\infty}\frac{Rdr}{r\sqrt{r^2-R^2}}=\frac{\pi}{2}, \\
  &\int_{R}^{\infty} \frac{M(R^2+rR+r^2)dr}{r^2(r+R)\sqrt{r^2-R^2}}=\frac{2M}{R},\\
 &\int_{R}^{\infty} \frac{3M^2(R^2+rR+r^2)^2dr}{2r^3R(r+R)^2\sqrt{r^2-R^2}}=\frac{M^2}{8R^2}(15\pi-16), \\
 &\int_{R}^{\infty} \frac{2Jdr}{R(r+R)\sqrt{r^2-R^2}}=\frac{2J}{R^2},\\
 &\int_{R}^{\infty} \frac{2MJ(5R^2+4rR+2r^2)dr}{rR^2(r+R)^2\sqrt{r^2-R^2}}=\frac{MJ}{R^3}(5\pi-8),\\
& \int_{R}^{\infty} \frac{J^2(R^2+rR+r^2)dr}{Mr^2R^2(r+R)\sqrt{r^2-R^2}}=\frac{2J^2}{MR^3},\\
& \int_{R}^{\infty} \frac{5M^3(R^2+rR+r^2)^3dr}{2r^4R^2(r+R)^3\sqrt{r^2-R^2}}=\frac{M^3}{R^3}\left(\frac{61}{3}-\frac{15\pi}{4}\right).
\end{align}
It is also seen that unlike in the Kerr metric, term proportional to  $\frac{a^2}{R^2}$ (originally $\frac{J^2}{M^2R^2}$ in the HT metric), which drops out due to the integration in the Kerr metric, does not occur at all in the HT metric.

\section{Deflection angle as a function of impact parameter}\label{App_b}
As was already derived above (see Eq.~(\ref{def_b})), impact parameter $b=\frac{p_{\varphi}}{\omega_{0}}$ can be expressed as a function of $R$ in terms
\begin{equation}
b=\frac{P(R)}{A(R)}\pm h(R)=\frac{P(R)}{A(R)}\pm n\sqrt{\frac{C(R)}{A(R)}+\frac{P^2(R)}{A^2(R)}}.
\end{equation}
Although there is a sign ambiguity, in Ref.~\cite{edery06} it was discussed in detail that without loss of generality, it is possible to choose only the positive solution. This is applied further. In accordance with the previous results, let us present the expressions up to the order $\propto\frac{M^3}{R^3}$.

At first, let us focus again on a vacuum case, i.e., set $n=1$. In that case, for the Kerr metric one gets
\begin{equation}\label{R_b_Kerr}
  b=\frac{R}{\sqrt{1-\frac{2M}{R}-\frac{a^2}{R^2}+\frac{4Ma}{R^2}-\frac{4M^2a}{R^3}}}.
\end{equation}
The last term is not applied at the given approximation (up to the third order); it is given only for completeness.
Let us further express $R$ as a function of $b$. This can straightforwardly be performed from the previous formula, and it leads to the result
\begin{equation}
  R=b\left(1-\frac{M}{b}-\frac{a^2}{2b^2}+\frac{2Ma}{b^2}-\frac{3M^2}{2b^2}\right).
\end{equation}
Only the terms relevant in the assumed approximation are given. First two terms correspond to the well known relation between $R$ and $b$, as was already noticed, e.g., in Ref.~\cite{edery06}. When the last relation is substituted into (\ref{alpha_BL}), we find
\begin{align}\label{aplha_b_Kerr}
\alpha_{BL}(b)=&\frac{4M}{b}+\frac{15\pi}{4}\frac{M^2}{b^2} \mp\frac{4Ma}{b^2}\mp\frac{M^2a}{b^3}(10\pi-8)\\
&-\frac{8M^2a}{b^3}+\frac{4Ma^2}{b^3}+\frac{128}{3}\frac{M^3}{b^3}.\nonumber
\end{align}

In the case of the HT metric when $K=0$, the relation between $R$ and $b$ takes the form
\begin{equation}
  b=\frac{-\frac{2Ma}{R}+\sqrt{R^2\left(1-\frac{2M}{R}\right)}}{\left(1-\frac{2M}{R}\right)\left(1-\frac{Ma^2}{R^3}\right)}.
\end{equation}
This expression can be rewritten in the form analogous to (\ref{R_b_Kerr}):
\begin{equation}
  b=\frac{R}{\sqrt{1-\frac{2M}{R}+\frac{4Ma}{R^2}-\frac{4M^2a}{R^3}-\frac{2Ma^2}{R^3}}},
\end{equation}
which gives (up to relevant terms)
\begin{equation}
  R=b\left(1-\frac{M}{b}+\frac{2Ma}{b^2}-\frac{3M^2}{2b^2}\right).
\end{equation}
This relation can again be substituted to the deflection angle formula obtained for the HT metric, Eq.~(\ref{alpha_HT0}). Hence, one gets
\begin{align}\label{aplha_b_HT}
\alpha_{HT0}(b)=&\frac{4M}{b}+\frac{15\pi}{4}\frac{M^2}{b^2} \mp\frac{4Ma}{b^2}\mp\frac{M^2a}{b^3}(10\pi-8)\\
&-\frac{8M^2a}{b^3}+\frac{4Ma^2}{b^3}+\frac{128}{3}\frac{M^3}{b^3}.\nonumber
\end{align}
As it can be seen from the expressions derived for $\alpha_{BL}(b)$ in (\ref{aplha_b_Kerr}) and $\alpha_{HT0}(b)$ in (\ref{aplha_b_HT}), these two relations are identical. It is thus seen that since the impact parameter is a coordinate independent variable (ratio of the constants of motion), in the weak field approximation when $K=0$, these two deflection angles coincide, as expected.

Let us further discuss the case of HT metric when $K\ne0$. However, at given approximation, i.e., up to the third order in $\frac{M}{R}$, and considering that the terms $\sim K$ are $\propto \frac{M^3}{R^3}$ and $\propto \frac{M^4}{R^4}$, respectively, these terms remain the same as in the previous case, only with substitution $R\rightarrow b$.

Let us now consider plasma, given in the same way as before, i.e., for refractive index it holds
\begin{equation}
n^2=1-\frac{\omega_{p}^2(r)}{\omega^2(r)}=1-\frac{\omega_{p}^2(r)}{\omega_0^2}A(r).
\end{equation}
This term also appears in the relation for the impact parameter, and it leads to an additional term which yields as
\begin{align}
b\approx&\frac{P(R)}{A(R)}\pm \frac{\sqrt{A(R)C(R)+P^2(R)}}{A(R)} \\
&\mp\frac{\omega_{p}^2(R)}{2\omega_0^2}\sqrt{A(R)C(R)+P^2(R)}\nonumber\\
=&b_0+ b_{pl}.\nonumber
\end{align}
In the Kerr metric the plasma part is given by
\begin{equation}
  b_{pl}=\frac{\omega_{p}^2(R)R}{2\omega_0^2}\left(1-\frac{M}{R}+\frac{a^2}{2R^2}-\frac{M^2}{2R^2}+\frac{Ma^2}{2R^3}-\frac{M^3}{2R^3}\right),
\end{equation}
demonstrating that when only the first plasma term in the deflection angle is applied, this expression can be neglected completely. In this text, it is mentioned only for completeness.

Terms which remain to be expressed as functions of $b$ instead of $R$ are still the plasma term $\alpha_{refr}$ and the combined term $\alpha_{refrHT}$. The recalculation of $\alpha_{refr}$ as a function of $b$ was already derived by \citet{bisnovatyi15}, and it yields
\begin{equation}
\alpha_{refr}(b)=\frac{\mathcal{C}_e}{\omega_{0}^2}\int_{0}^{\infty}\frac{\partial N}{\partial b}dz.
 \end{equation}
Because term $\alpha_{refrHT}$ was treated in the similar manner, let us briefly repeat the procedure of the $\alpha_{refr}(b)$ derivation as performed in Ref.~\cite{bisnovatyi15}.

Let us begin with a simple substitution in the form
\begin{gather}\label{alpha_refr_b}
\alpha_{refr}(R)=\frac{R}{\omega_{0}^2}\int_{R}^{\infty}\frac{r(\omega_{p}^2(r)-\omega_{p}^2(R))dr}{(r^2-R^2)^{3/2}}=\\\begin{vmatrix}
sub. & d\left(\frac{1}{\sqrt{r^2-R^2}}\right) =& \frac{-r}{(r^2-R^2)^{3/2}}
\end{vmatrix}=\nonumber\\
-\frac{R}{\omega_{0}^2}\int_{R}^{\infty}(\omega_{p}^2(r)-\omega_{p}^2(R))d\left(\frac{1}{\sqrt{r^2-R^2}}\right)=\nonumber\\
\left.-\frac{R(\omega_{p}^2(r)-\omega_{p}^2(R))}{\omega_{0}^2\sqrt{r^2-R^2}}\right|^{\infty}_R+
\frac{R}{\omega_{0}^2}\int_{R}^{\infty}\frac{1}{\sqrt{r^2-R^2}}\frac{d\omega_{p}^2(r)}{dr}dr.\nonumber
\end{gather}
The first term drops out because for $r\rightarrow\infty$ it obviously goes to zero, while when $r\rightarrow R$ one can write
\begin{gather}
\left.\frac{R(\omega_{p}^2(r)-\omega_{p}^2(R))}{\omega_{0}^2\sqrt{r^2-R^2}}\right|_{r\rightarrow R}\approx \left.\frac{R}{\omega_{0}^2\sqrt{2R(r-R)}}\right.\\
\left.\times\left(\omega_{p}^2(R)+\left.\frac{d\omega_{p}^2(r)}{dr}\right|_{r=R}(r-R)-\omega_{p}^2(R)\right)\right|_{r\rightarrow R},\nonumber
\end{gather}
which eventually gives zero as well.

The second term of Eq.~(\ref{alpha_refr_b}) can be further treated when considering that $\omega_{p}^2(r)=\mathcal{C}_eN(r)$, where $\mathcal{C}_e$ is a constant factor given by combination of physical constants and $N(r)$ denotes the plasma number density which is given by the concrete plasma distribution. Thus one can further write
\begin{gather}
\frac{R}{\omega_{0}^2}\int_{R}^{\infty}\frac{1}{\sqrt{r^2-R^2}}\frac{d\omega_{p}^2(r)}{dr}dr =\frac{R\mathcal{C}_e}{\omega_{0}^2}\int_{R}^{\infty}\frac{1}{\sqrt{r^2-R^2}}\frac{dN}{dr}dr\nonumber\\
=\begin{vmatrix}
sub. & r^2&=&b^2+z^2\\
& dr&=&\frac{zdz}{r}
\end{vmatrix}
\begin{vmatrix}
 R=b,& r^2-R^2=z^2\\
\frac{\partial N}{\partial b}&=\frac{\partial r}{\partial b}\frac{dN}{dr}=\frac{b}{r}\frac{dN}{dr}
\end{vmatrix}\nonumber\\
=\frac{\mathcal{C}_e}{\omega_{0}^2}\int_{0}^{\infty}\frac{\partial N}{\partial b} dz\equiv\alpha_{refr}(b).
\end{gather}
In this calculation it was assumed that $b=R$, but as was already discussed above, it is a sufficient approximation for the calculated term. The transformation was performed under the assumption that the unperturbed light trajectory is a straight
line parallel to the $z$-axis with impact parameter $b$, and it also holds $r=\sqrt{b^2+z^2}$.

Let us now perform the similar calculation for the term $\alpha_{refrHT}$. At first, it is desired to rewrite the term to be similar to the previous case, and to perform the same substitution as before. This leads to
\begin{gather}\label{alpha_refrHT_dr}
 \alpha_{refrHT}(R)=\frac{8KM^3}{5\omega_{0}^2}\int_{R}^{\infty}\frac{\left(R^3\omega_{p}^2(r)-r^3\omega_{p}^2(R)\right)dr}{r^2R^2(r^2-R^2)^{3/2}}\\
 =\frac{8KM^3}{5\omega_{0}^2}\int_{R}^{\infty}\left(\frac{\omega_{p}^2(r)}{r^3}-\frac{\omega_{p}^2(R)}{R^3}\right)\frac{rRdr}{(r^2-R^2)^{3/2}}\nonumber\\\begin{vmatrix}
sub. & d\left(\frac{1}{\sqrt{r^2-R^2}}\right) =& \frac{-r}{(r^2-R^2)^{3/2}}
\end{vmatrix}\nonumber\\
=-\frac{8KM^3R}{5\omega_{0}^2}\int_{R}^{\infty}\left(\frac{\omega_{p}^2(r)}{r^3}-\frac{\omega_{p}^2(R)}{R^3}\right)d\left(\frac{1}{\sqrt{r^2-R^2}}\right)\nonumber\\
=\left.-\frac{8KM^3R}{5\omega_{0}^2\sqrt{r^2-R^2}}\left(\frac{\omega_{p}^2(r)}{r^3}-\frac{\omega_{p}^2(R)}{R^3}\right)\right|^{\infty}_R\nonumber\\
+\frac{8KM^3R}{5\omega_{0}^2}\int_{R}^{\infty}\frac{1}{\sqrt{r^2-R^2}}\frac{d}{dr}\left(\frac{\omega_{p}^2(r)}{r^3}\right)dr.\nonumber
 \end{gather}
The same arguments as in the previous case can be applied to show that the first term drops out in the given limits. The second term can be treated as before, i.e.,
\begin{gather}
\frac{8KM^3R}{5\omega_{0}^2}\int_{R}^{\infty}\frac{1}{\sqrt{r^2-R^2}}\frac{d}{dr}\left(\frac{\omega_{p}^2(r)}{r^3}\right)dr=\\
\frac{8KM^3R\mathcal{C}_e}{5\omega_{0}^2}\int_{R}^{\infty}\frac{1}{\sqrt{r^2-R^2}}\frac{d}{dr}\left(\frac{N}{r^3}\right)dr=\nonumber\\
\begin{vmatrix}
\begin{array}{rcl}
sub. &r^2=b^2+z^2, \\
&dr=\frac{zdz}{r},\\
&R=b, r^2-R^2=z^2,
\end{array}
\frac{d}{dr}\left(\frac{N}{r^3}\right)=\frac{1}{r^2}\left(\frac{1}{b}\frac{\partial N}{\partial b}-\frac{3N}{r^2}\right)
\end{vmatrix}\nonumber\\
=\frac{8KM^3\mathcal{C}_e}{5\omega_{0}^2}\int_{0}^{\infty}\left(\frac{1}{b}\frac{\partial N}{\partial b}-\frac{3N}{b^2+z^2}\right) \frac{bdz}{(b^2+z^2)^{3/2}}\nonumber\\
\equiv\alpha_{refrHT}(b).\nonumber
\end{gather}
And that brings all relevant terms to be functions of $b$ instead of $R$.

For completeness, let us briefly express the relation between $b$ and $R$ also for other metrics discussed above. The relation for the impact parameter as a function of $R$ for a spherically symmetric metric can be obtained from the equation defined above when setting $P(r)=0$. This gives \footnote{The formula was recently nicely discussed in Ref.~\cite{PerlickTsupko22}.}
 \begin{equation}
b=\sqrt{\frac{C(R)}{A(R)}n^2}.
\end{equation}
For the Schwarzschild metric the relation is
\begin{equation}
  R=b\left(1-\frac{M}{b}-\frac{3M^2}{2b^2}\right),
\end{equation}
which follows from the result obtained in the Kerr metric when $a=0$. Moreover, because the corrections stemming from the quadrupole moment in the ER metric are at least of the third order, in the given approximation the relation introduced above holds also for the ER metric.

In the case of the $q$-metric the transformation is given by
\begin{equation}
  R=b\left(1-\frac{\mathcal{M}_q}{b}(1+2q)-\frac{\mathcal{M}_q^2}{2b^2}(3+4q(q+2))\right).
\end{equation}
It can be seen that for $q=0$ the formula takes the form valid for the Schwarzschild metric.

\bibliography{bibliography}

\end{document}